\documentclass[acmsmall, screen]{acmart}
\usepackage{amsthm, bm}
\usepackage[ruled,linesnumbered]{algorithm2e}
\usepackage{dsfont}
\usepackage{multirow}
\usepackage[normalem]{ulem}
\usepackage[utf8]{inputenc}
\usepackage[finnish, english]{babel}
\useunder{\uline}{\ul}{}
\newtheorem{assumption}{Assumption}
\AtBeginDocument{%
  \providecommand\BibTeX{{%
    \normalfont B\kern-0.5em{\scshape i\kern-0.25em b}\kern-0.8em\TeX}}}

\begin{document}

\title{Tail-Learning: Adaptive Learning Method for Mitigating Tail Latency in Autonomous Edge Systems}

\author{Cheng Zhang}
\email{coolzc@zju.edu.cn}
\affiliation{%
  \institution{Zhejiang University}
  \city{Hangzhou}
  \state{Zhejiang}
  \country{China}
}

\author{Yinuo Deng}
\email{yinuo@zju.edu.cn}
\affiliation{%
  \institution{Zhejiang University}
  \city{Hangzhou}
  \state{Zhejiang}
  \country{China}
  }

\author{Hailiang Zhao}
\authornote{Corresponding author.}
\email{hliangzhao@zju.edu.cn}
\affiliation{%
  \institution{Zhejiang University}
  \city{Hangzhou}
  \state{Zhejiang}
  \country{China}
}

\author{Tianlv Chen}
\email{phchtl@zju.edu.cn}
\affiliation{%
  \institution{Zhejiang University}
  \city{Hangzhou}
  \state{Zhejiang}
  \country{China}
 }

\author{Shuiguang Deng}
\email{dengsg@zju.edu.cn}
\affiliation{%
  \institution{Zhejiang University}
  \city{Hangzhou}
  \state{Zhejiang}
  \country{China}
  }
\renewcommand{\shortauthors}{Zhang et al.}

% abstract
%================================================================
\begin{abstract}
In the realm of edge computing, the increasing demand for high Quality of Service (QoS), particularly in dynamic multimedia streaming applications (e.g., Augmented Reality/Virtual Reality and online gaming), has prompted the need for effective solutions.
Nevertheless, adopting an edge paradigm grounded in distributed computing has exacerbated the issue of tail latency.
Given a limited variety of multimedia services supported by edge servers and the dynamic nature of user requests, employing traditional queuing methods to model tail latency in distributed edge computing is challenging, substantially exacerbating head-of-line (HoL) blocking.
In response to this challenge, we have developed a learning-based scheduling method to mitigate the overall tail latency, which adaptively selects appropriate edge servers for execution as incoming distributed tasks vary with unknown size. To optimize the utilization of the edge computing paradigm, we leverage Laplace transform techniques to theoretically derive an upper bound for the response time of edge servers.
Subsequently, we integrate this upper bound into reinforcement learning to facilitate tail learning and enable informed decisions for autonomous distributed scheduling.
The experiment results demonstrate the efficiency in reducing tail latency compared to existing methods.
\end{abstract}




\maketitle

%================================================================
% Introduction
%================================================================
\section{Introduction}
Edge computing is a computation paradigm that executes user tasks locally with nearby servers. 
In the context of \textit{distributed edge computation}, users can concurrently select multiple edge servers for parallel task execution \cite{zhang2022aggcast}. 
The considerations in distributed edge computation extend beyond the array of services supported by edge servers to encompass the concept of \textit{composed latency} \cite{yue2021todg, wang2019effective}. 
For example, when aggregating data from cross-edge users or end devices for analysis, it is essential to factor in the latency associated with uplink/downlink transmissions, collaboration latency across edge servers, and the intrinsic data processing time \cite{jia2018optimizing}. 
Minimizing composed latency emerges as a critical imperative for improving user experiences \cite{bouet2018mobile}.
\begin{figure}[htbp]
\centering
    \includegraphics[width=0.7\linewidth]{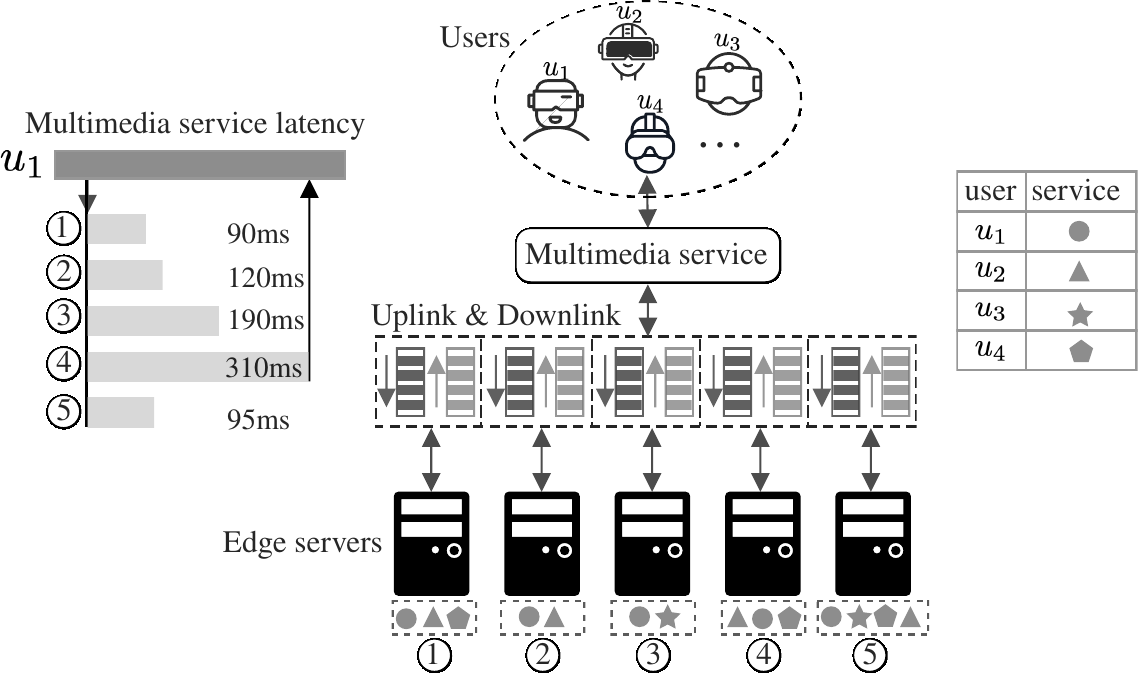} 	
    \caption{In the considered distributed edge computation scenario, users initiate diverse multimedia service requests randomly, and each edge server is designed to support a specific range of rendering/transcoding service request types. 
    The system enhances execution efficiency by executing service requests in parallel, leveraging a group of edge servers with varying degrees of parallelism. 
    For example, the service request initiated by user $u_1$ can be processed by all the five edge servers by launching a sub-task on each edge server, respectively. 
    The left side of the figure displays a request latency diagram, portraying the response time of each parallel sub-task as the cumulative sum of uplink, execution, and downlink delays. The maximum response time among these parallel sub-tasks, reaching $310$ms, is the latency experienced by $u_1$.}
    \label{fig_intro}
\end{figure}
Besides, given the resource-intensive nature of multimedia applications (e.g., AR/VR and online gaming), leveraging the substantial computing resources offered by edge servers can markedly enhance the user experience on various devices \cite{pal2023edgemart, 8560111, tan2020uav, mehrabi2021multi, chen2019collaborative, 8818403, 8542668}. 
Edge servers provide the advantage of delivering services with lower latency to end users compared to cloud servers. 
Edge computing can shift rendering tasks from end-user devices to geographically proximate edge servers. 
In the context of AR/VR applications assisted by distributed edge computing, the partitioned rendering workload is essential for ensuring a seamless and immersive user experience. 
By utilizing distributed edge computing, rendering/transcoding workloads in AR/VR can be partitioned and executed across various distributed computing resources \cite{alshahrani2020efficient, liu2019reliability}. 
Edge servers collaborate to synchronize and composite different views, creating a shared virtual experience \cite{zhang2019rendering}.
Reduced rendering latency leads to a more responsive and immersive AR/VR experience. 
Numerous studies have demonstrated that distributing rendering/transcoding workloads using edge servers enhances the efficiency and responsiveness of rendering in augmented and virtual reality applications \cite{tan2020uav, mehrabi2021multi, chen2019collaborative, alshahrani2020efficient, liu2019reliability, zhang2019rendering}.

However, recent research has predominantly focused on achieving resource-efficient delay minimization, as evidenced by works such as \cite{zhang2022aggcast, 8818444, li2022efficient, yue2021todg, 8818403, wang2019effective, jia2018optimizing, bouet2018mobile, 9154603}. 
These endeavors overlook the crucial aspect of \textit{tail latency} effects. 
% Tail latency is the delay in response that exceeds a specified threshold, commonly measured using a specific percentile to capture, e.g., the $99$-th ($p99$) is the threshold below which $99\%$ of service latency resides \cite{chen2021achieving, mirhosseini2020q, mirhosseini2019express}. 
Tail latency refers to the delay in response that exceeds a predefined threshold, typically measured using a specific percentile to capture the extreme values. 
For instance, the $99$-th percentile ($p99$) represents the threshold below which $99\%$ of service latency is observed \cite{chen2021achieving, mirhosseini2020q, mirhosseini2019express}.
Considerable long tail delays in certain tasks can severely impede the processing of subsequent incoming tasks, leading to a decrease in QoS.
In edge computing, long tail latency, usually shortened as tail latency, can result in significant head-of-line (HoL) blocking problems \cite{chen2021achieving}. 
Moreover, tail latency typically emerges in response to requests from the highest-paying users, generating the most computation usage. 
Improving these users' experiences is particularly imperative, given the inverse correlation between service response and revenue, and the significant impact of increased response time on user experience \cite{hussain2022cloud, kleppmann2017designing}. 
However, tail latency is exacerbated in distributed edge computation. 
Fig. \ref{fig_intro} gives an example. 
In the considered scenario, tasks with different service requests can run concurrently on multiple servers. 
For instance, when user $u_1$'s service request is executed in parallel across all five edge servers, the response time is determined by the longest sub-task time, resulting in a service response time of 310ms. 
This situation is referred to as the \textit{amplification of tail latency} \cite{dean2013tail}. 
Whenever a user's service request necessitates multiple edge servers to run in parallel, even with only a small fraction of edge servers displaying slow response times during parallel execution, the probability of encountering slow execution rises. 
This results in a higher proportion of service requests experiencing tail latency.

Effectively addressing the formidable challenge of mitigating tail latency amplification in distributed edge computation demands developing more efficient strategies.
First, the intrinsic challenge of reducing tail latency for online service requests stems from the stochastic nature of user requests in the dynamic edge environment.
Existing approaches, such as the weighted max-min algorithm \cite{xie2022qos}, block coordinate descent algorithm \cite{zhang2019rendering}, and greedy-based algorithm \cite{mehrabi2021multi, kumar2020fast}, often use offline or deterministic methods, making them ill-suited to handle the online task-arriving scenario.
Second, optimization strategies grapple with significant complexity due to constraints like limited and heterogeneous computing resources and imbalanced service deployment across edge servers. 
Mathematical modeling of tail latency in this complex landscape becomes particularly intricate, leading to non-convex formulations. 
Some researchers resort to machine learning or heuristic methods to model tail latency, such as the learning for prediction method \cite{rahman2019predicting}, estimation-based load-aware scheme \cite{xie2018cutting}, and the Size-Interval-Task-Assignment method \cite{mirhosseini2020q}. 
However, these approaches may be unsuitable for edge computing scenarios and require more theoretical analysis of tail latency to improve the algorithm efficiency. 
Thus, more readily available solvers are imperative to determine optimal solutions.

To address these challenges, we introduce a reinforcement learning-based algorithm called \textit{Tail-learning} in this paper. We employ a probabilistic scheduling model, as utilized by \cite{aggarwal2017taming, xiang2014joint}, to represent the decision variable as the probability of assigning a specific service request to a particular edge server. Although the decision variables in the optimization problem remain non-convex, leveraging queuing theory allows us to model the service response time of the edge network. Specifically, we adopt a tandem queue network to establish an analytical upper bound for the tail latency probability of edge servers \cite{harchol2013performance, kleinrock1975queueing, liang2023model, wang2019effective}. 
This approach demonstrates that the optimization problem becomes convex for the Laplace transform parameters. 
Consequently, we employ the Laplace Transform of server response time \cite{kleinrock1975queueing}, the upper bound of server tail probability, and their first and second-order mathematical forms as the state of the reinforcement learning algorithm. 
Additionally, to mitigate the level of \textit{HoL} blocking, we incorporate the congestion of queue length as a penalty in the reward function, where the tail latency is quantified by the threshold delay parameter. 
Moreover, in the autonomous edge system, our proposed design significantly reduces the action space by employing a framework that includes the \textit{Shared Representation Layer} and multiple decision-making \textit{agents}.
% Furthermore, to reduce the \textit{HoL} blocking level, we adopt the congestion of queue length as a penalty in the reward function, and the tail latency measured the threshold delay.   
% Besides, in the autonomous edge system, our proposed design can reduce the action space tremendously by using the framework consisting of \textit{Shared Representation Layer} and multiple decision-making agents.  
This novel approach enables us to learn optimal decisions, effectively reducing the tail latency of the edge network. 
Our method provides a rigorous and systematic performance analysis and modeling of the edge network, offering a comprehensive exploration of hidden metrics associated with tail latency. 
Notably, the states in the reinforcement learning framework prove effective and efficient in characterizing the response time of edge servers. 
Therefore, the decision-making algorithm we propose continuously adjusts strategies to adapt to environmental changes and optimize decision-making performance through dynamic interaction with the environment. 
This \textit{adaptive} property enables our proposed \textit{Tail-learning} algorithm to perform well in edge computing environments where service requests change dynamically.

The main contributions are summarized as follows.
\begin{itemize}
    \item 
    We formulate the response time dynamics of distributed edge servers with a robust queuing theory model. 
    We not only model the response time but also establish a theoretical bound for the tail-latency probability. 
    This foundational aspect provides a comprehensive understanding of the latency landscape in distributed edge computation.
    \item 
    We design a reinforcement learning-based algorithm \textit{Tail-learning} that transforms the online non-convex problem associated with tail latency into a \textit{learnable} online decision-making problem. 
    By bridging the gap between theory and practical implementation, \textit{Tail-learning} offers a tangible and effective solution to address the challenges posed by tail latency in distributed edge environments. 
    Compared with reinforcement learning modeled by one single agent, our method reduces the action encoding dimension from \(|B_i|^I\) to \(\sum_{i=0}^I{|B_i|}\).
    \item 
    The superiority of the proposed algorithm is demonstrated through extensive experimentation, showcasing its efficacy in addressing the intricate challenges posed by tail latency in distributed edge computation. 
    At the $p99.9$ percentile, the experimental results of our proposed algorithm reveal only $60.78\%$ of the long-tail delay compared to the optimal benchmarking policy.
\end{itemize}

The rest of the paper is organized as follows. Section \ref{sec_related_work} discusses related work on distributed edge computation systems. Section \ref{sec_prob_formulation} details the system modeling for the distributed edge computation system based on queuing theory. Section \ref{sec_tail_analysis} analyzes the tail-latency probability bound through rigorous mathematical proof. Section \ref{sec_tail_learning} introduces the designed tail-learning algorithm based on reinforcement learning. Section \ref{sec_evaluation} presents an experimental analysis of the tail-learning algorithm. Section \ref{sec_conculsion} outlines the conclusions and future work.
%================================================================
% Related Work
%================================================================
\section{Related Work}\label{sec_related_work}
%================================================================
% Modeling of Tail-latency Optimization
%================================================================
We review relevant literature, focusing on two aspects: distributed modeling and algorithm design.
\subsection{Distributed Modeling}
Some research focuses on edge networks' computing or bandwidth resource limitations and designs strategies to allocate user service requests to different servers for computation. 
For instance, Zhang et al. \cite{zhang2022aggcast} propose the optimization goal of minimizing bandwidth resources based on various services . 
In the case of dynamic changes in edge network resources, Yue et al. \cite{yue2021todg} design a service request framework to ensure that the latency of the requested task is guaranteed within the specified time. 
Toczé et al. \cite{tocze2019performance} focus on the latency and throughput by offloading the video streaming data to the edge server through communication links.
In the context of 5G, Tan et al. \cite{tan2020uav} present a sophisticated edge computing framework employing unmanned aerial vehicles (UAVs) to provide energy-efficient AR services with minimal delay constraints.
While these studies consider latency in the context of resource and service requests in edge environments, none explicitly addresses the optimization of distributed computing patterns in such environments. 
Mirhosseini et al. \cite{mirhosseini2020q} propose a queuing theory method based on an express lane for short microservices tasks to reduce tail delay. 
Additionally, Wang et al. \cite{wang2019effective} employ the tandem queue method to model the transmission link queue in edge scenarios concurrently with the edge server queue to ensure task delay requirements. 
While both works adopt queuing theory methods for modeling distributed computing, they do not provide a solution for optimizing the task aggregation latency in distributed edge computing. 
Authors in \cite{jia2018optimizing,li2022efficient, liu2017stepwise} present an optimization method for aggregating different data services on several edge servers in an edge environment. 
Zhang et al. \cite{zhang2019rendering} focus on optimizing the video rendering and encoding of edge-based Mobile Augmented Reality (MAR) services, specifically addressing multi-party applications formulating an efficient task assignment approach, ultimately enhancing QoS and edge computing efficiency.
They propose an optimal data segmentation and aggregation strategy to minimize task delay. 
However, the aggregation latency metric in these works lacks consideration of long-tail latency.
%================================================================
% Algorithms for Tail-latency Optimization
%================================================================
\subsection{Algorithm Design}
Bouet et al. \cite{bouet2018mobile} employ a mixed-integer optimization method to address the optimal allocation problem for limited resources, minimizing user task delay within the edge network. 
Alshahrani et al. \cite{alshahrani2020efficient} introduce an efficient multi-player with multi-task computation offloading model for VR applications in mobile edge-cloud computing, formulated as an integer optimization problem for solving optimal decision-making. 
Xie et al. \cite{xie2022qos} address the problem of scheduling multiple 3D multimedia rendering tasks with diverse QoS fairness requirements in edge computing environments by using the water filling algorithm to solve a max-min utility problem.
However, this deterministic offline algorithm overlooks the online arrival of unpredictable user tasks. 
The uncertainty introduced by online tasks makes solving online task problems more challenging than their offline counterparts, necessitating the development of more adaptive methods for online task problem-solving.
Mehrabi et al. \cite{mehrabi2021multi} utilize multi-access edge computing to optimize the tradeoff between average video quality and delivery latency in remote-rendered interactive VR. They formulate the problem as a mixed-integer nonlinear programming model and design an online greedy algorithm to solve it.
Furthermore, Xie et al. \cite{xie2018cutting} address the challenge of tail latency in networks, proposing a load-aware selection scheme based on a heuristic method, demonstrating substantial reductions in long-tail latency. 
Kumar et al. \cite{kumar2020fast} classify server resources into Fast and Frugal, and then implement different heuristic policies to allocate tasks to servers. 
While the above algorithms focus on optimization methods for distributed computing patterns in edge environments, simple heuristic techniques may fail to deliver the optimal decisions for distributed edge computation. 
They lack a theoretical analysis of the aggregate tail latency of each server or service. 
Nishtala et al. \cite{nishtala2020twig} propose a reinforcement learning-based algorithm to characterize tail latency by monitoring task quality of service, thereby learning the optimal strategy to reduce tail latency.
Raeis et al. \cite{raeis2021queue} introduce a reinforcement learning-based service-rate controller to minimize the end-to-end delay in tandem service systems, offering explicit guarantees on QoS constraints.
While they employ reinforcement learning methods to minimize the network system's latency, their studies lack an in-depth exploration of system theoretical modeling.
The algorithm we propose establishes a tail-upper bound through queuing theory analysis and combines this theoretical bound with corresponding first-order and second-order expressions. 
Our proposed reinforcement learning approach enables the design of an optimization algorithm that can be learned online, thereby significantly reducing long-tail latency.
\begin{table}[]
\centering
\begin{tabular}{|c|cccccc|}
\hline
\multirow{2}{*}{Reference} &\multicolumn{3}{c|}{Distributed Modeling} & \multicolumn{3}{c|}{Algorithm Design} \\\cline{2-7}
& \begin{tabular}[c]{@{}c@{}} Minimal\\ latency \end{tabular} 
& \begin{tabular}[c]{@{}c@{}}Distributed\\manner\end{tabular} 
& \begin{tabular}[c]{@{}c@{}}Aggregation\\latency\end{tabular} 
& \begin{tabular}[c]{@{}c@{}}Online\\algorithm\end{tabular} 
& \begin{tabular}[c]{@{}c@{}}Adaptive\\learning\end{tabular} 
& \begin{tabular}[c]{@{}c@{}}Tail\\upper-bound\end{tabular} \\\hline
\multicolumn{1}{|c|}{\cite{zhang2022aggcast,yue2021todg, tocze2019performance, tan2020uav}}                                  & \multicolumn{1}{c|}{\checkmark}                           & \multicolumn{1}{c|}{-}                                       & \multicolumn{1}{c|}{-}                                        & \multicolumn{1}{c|}{-}                                     & \multicolumn{1}{c|}{-}                                      & -                                                          \\ \hline
\multicolumn{1}{|c|}{\cite{mirhosseini2020q,wang2019effective,bouet2018mobile,alshahrani2020efficient,xie2018cutting}} & \multicolumn{1}{c|}{\checkmark}                           & \multicolumn{1}{c|}{\checkmark}                              & \multicolumn{1}{c|}{-}                                        & \multicolumn{1}{c|}{-}                                     & \multicolumn{1}{c|}{-}                                      & -                                                          \\ \hline
\multicolumn{1}{|c|}{\cite{jia2018optimizing,li2022efficient,liu2017stepwise,zhang2019rendering}}                             & \multicolumn{1}{c|}{\checkmark}                           & \multicolumn{1}{c|}{\checkmark}                              & \multicolumn{1}{c|}{\checkmark}                               & \multicolumn{1}{c|}{-}                                     & \multicolumn{1}{c|}{-}                                      & -                                                          \\ \hline
\multicolumn{1}{|c|}{\cite{mehrabi2021multi,xie2018cutting,kumar2020fast}}                                   & \multicolumn{1}{c|}{\checkmark}                           & \multicolumn{1}{c|}{\checkmark}                              & \multicolumn{1}{c|}{-}                                        & \multicolumn{1}{c|}{\checkmark}                            & \multicolumn{1}{c|}{-}                                      & -                                                          \\ \hline
\multicolumn{1}{|c|}{\cite{nishtala2020twig, raeis2021queue}}                                                                               & \multicolumn{1}{c|}{\checkmark}                           & \multicolumn{1}{c|}{\checkmark}                              & \multicolumn{1}{c|}{-}                                        & \multicolumn{1}{c|}{\checkmark}                            & \multicolumn{1}{c|}{\checkmark}                             & -                                                          \\ \hline
\multicolumn{1}{|c|}{Ours}                                                                                                                         & \multicolumn{1}{c|}{\checkmark}                           & \multicolumn{1}{c|}{\checkmark}                              & \multicolumn{1}{c|}{\checkmark}                               & \multicolumn{1}{c|}{\checkmark}                            & \multicolumn{1}{c|}{\checkmark}                             & \checkmark                                                 \\ \hline
\end{tabular}
\caption{Comparison of different optimizations for latency minimization}
\label{tab:ref_constrast}
\end{table}
Table \ref{tab:ref_constrast} summarizes the references of the aforementioned tail-latency optimization studies.
%================================================================
% Problem Model and Formulation
%================================================================
\section{Problem Model and Formulation}\label{sec_prob_formulation}
%================================================================
% Motivation
%================================================================
\subsection{Motivation}
\begin{figure}[htbp]
    \centering
    \includegraphics[width=0.6\textwidth]{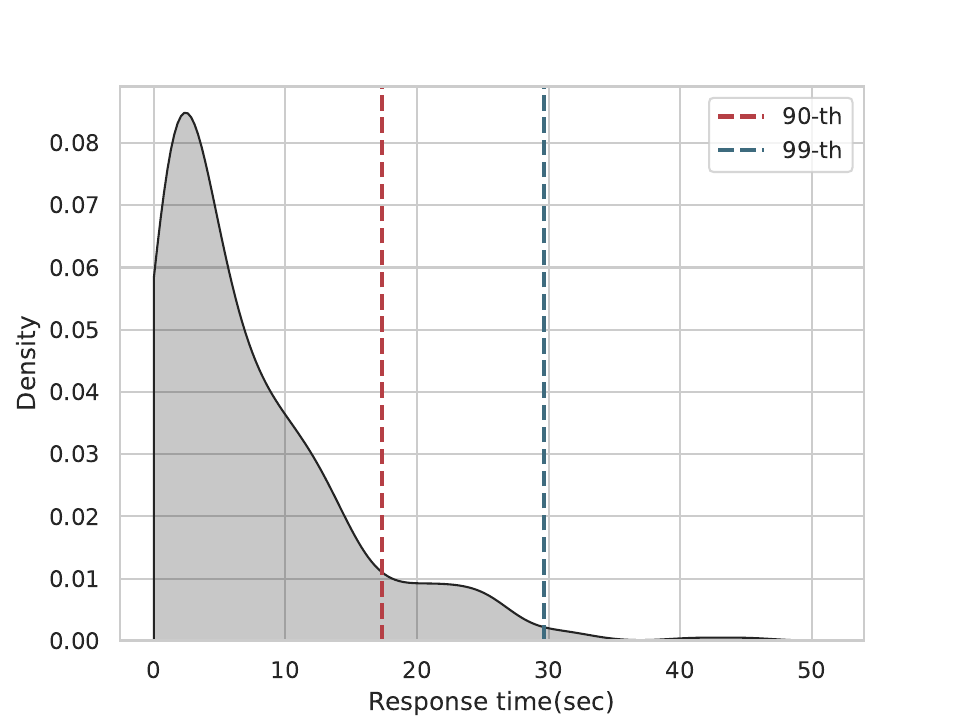} 	
    \caption{Probability Density of Response Time.}
    \label{fig_motivation}
\end{figure}
When streaming AR/VR and live gaming over the Internet, it is often imperative to adapt the original video file to a specific format, adjusting parameters like bitrate, encoding schemes, resolution, and file containers. This adaptation caters to diverse Internet connections and end-user device environments. With the evolution of wireless communications, the capability to perform video transcoding at the distributed edge has become feasible. To illustrate the amplification of tail latency, we instantiate a \textit{distributed video transcoding} task, converting M2TS Blu-ray videos into the Matroska format \cite{noe2007matroska}. We gather response time data from this distributed video transcoding task, presenting the probability density distribution diagram of response time in Figure \ref{fig_motivation}. The horizontal axis signifies the response time of distributed service requests, while the vertical axis depicts the probability of a specific response time. It reveals that the majority of service requests' response times concentrate within the header range of the horizontal axis. However, we can see that, at the $99$-th or $95$-th percentiles in the figure, several service requests experience considerably prolonged response times distributed in the tail range of the horizontal axis. These tail delays significantly impact user experiences. Our target is to mitigate the \textit{probability of tail latency} by devising effective algorithms. 

Intuitively, our goal is to push the $99$-th or $95$-th percentile indicator to the left as much as possible, thereby lowering the tail latency threshold. This reduction in tail latency is crucial for enhancing user experience and ensuring a smoother streaming service.
% When streaming video files over the Internet, it is necessary to convert the original video file into a specific format by altering bitrate, encoding algorithms, file containers, etc., to accommodate varying Internet connections and end-user device environments. With the development of wireless communications, it is possible to do video transcoding at the distributed edge. To show the amplification of tail latency, we construct a \textit{distributed video transcoding} task in which  the M2TS Blu-ray videos are transcoded into the Matroska format \cite{noe2007matroska}. We collect the response time data from the above distributed video transcoding task and present the probability density distribution diagram of response time in Fig. \ref{fig_motivation}. The horizontal axis denotes the response time of distributed service requests, while the vertical axis represents the probability of response time. It demonstrates that the most service requests' response time concentrates on the horizontal axis's header range. However, as indicated at the $99$-th or $90$-th percentiles in the figure, a small number of service requests exhibit significantly long response times distributed in the tail range of the horizontal axis. These tail delays significantly impact users' QoS. We aim to reduce the \textit{probability of tail latency} by designing effective algorithms. Intuitively, the aim is to shift the $99$-th or $95$-th percentile indicator to the left, lowering the tail latency threshold.
 \begin{table}[htbp]   
    \center
    \caption{\label{tab:summary}Summary of key notations.}   
    \begin{tabular}{ll}    
        \toprule
        {\textsf{\textbf{Notation}}}& {\textsf{\textbf{Description}}}\\
        \midrule
        $\mathcal{M}$ &Set of servers in the network\\
        $\mathcal{I}$ &Set of supplied services in the network\\
        $\Lambda_j$                 &Aggregated arrival rate to the $j$-th server\\
        $\mu_u^j, \mu_s^j, \mu_d^j$ &Service rate of the $j$-th uplink, server and downlink\\
        $T_j$         &Response time of the $j$-th tandem queueing \\
        $\hat{T}_j$         &\textit{Laplace Transform}s of $T_j$ \\
        $T_u^j, T_s^j, T_s^j$       &Response time of the $j$-th uplink, server and downlink\\
        $\hat{T}_u^j, \hat{T}_s^j$, $\hat{T}_d^j$ &\textit{Laplace Transform}s of $T_u^j$, $T_s^j$ and $T_d^j$ \\
        $B_i$                    &Set of candidate servers that can support service $i$  \\
        $\omega_{ij}$               &Decision of probability scheduling for service $i$ to server $j$\\
        $\gamma$                    &Tolerance level for the tail latency \\
        % $L_i$                       &Random variable of latency of service $i$ \\
        \bottomrule   
    \end{tabular}  
  \end{table}
%================================================================
% System Modeling
%================================================================
\subsection{System Modeling}
\subsubsection{Response Time}\label{sec_res_t}
\begin{figure}[htbp]
    \includegraphics[width=0.7\textwidth]{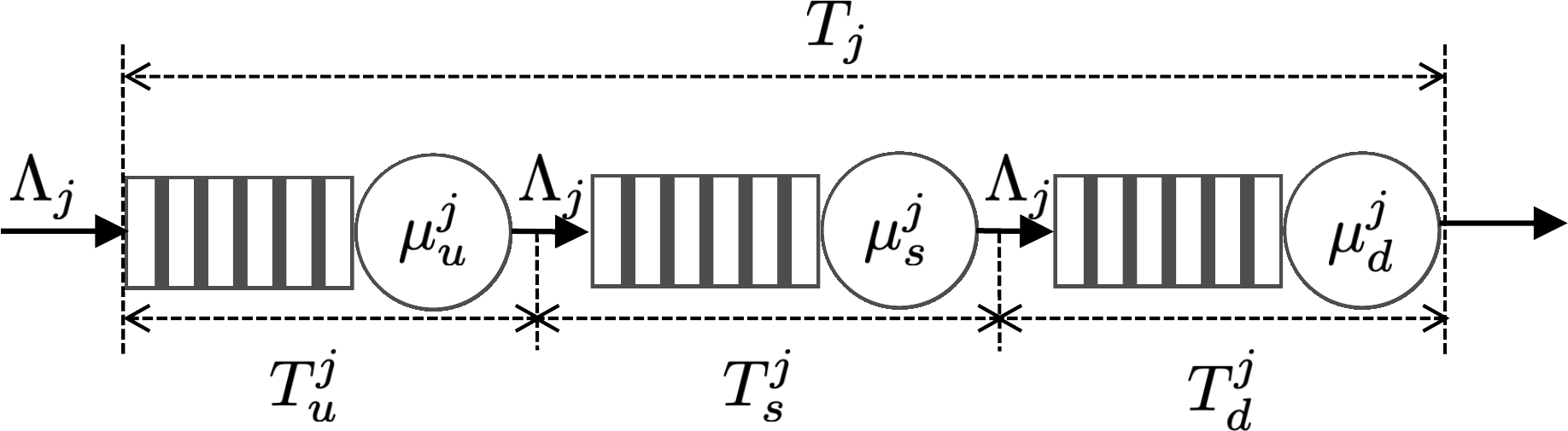} 	
    \caption{Tandem queuing network comprising uplink, edge server, and downlink.}
    \label{fig_system}
\end{figure}
We denote the set of server indices by $\mathcal{M} = \{1,2, \cdots, M\}$. 
Each server is equipped with corresponding uplink and downlink channels for transmitting service requests. 
The edge network provides services identified by $\mathcal{I} = \{1, 2, \cdots, I\}$. 
% Assign the index of each service request as $n$, and denote the response time of service request $n$ as $d_n$.

We focus on mitigating tail latency in the response time within a distributed edge network. 
The response time is defined as the interval between the departure of a service request from the user end and the subsequent return of the user result. 
We model each uplink, server, and downlink queue as an $M/M/1$ queue \cite{liu2017multiobjective, yuchong2019task}. 
In Figure \ref{fig_system}, the server, uplink, and downlink transmission links can be collectively considered a tandem network \cite{wang2019effective}.
For each $j$-th tandem queueing network of Figure \ref{fig_system}, we denote the response time by a random variable, i.e.,
\begin{equation}
    T_j = T_u^j + T_s^j + T_d^j,\label{formula_response_time}
\end{equation}
where $T_u^j, T_s^j$ and $T_d^j$ represent the random variables of latency for uplink $j$, server $j$ and downlink $j$, respectively. 
%================================================================
% Decisions for Probability Scheduling
%================================================================
\subsubsection{Decisions for Probability Scheduling}\label{sec_decision_Ai}
Because of unforeseen arrivals of user service requests, edge networks have to improve the schedules by taking into consideration uncertainty.
We employ tail-bound analysis to implement a probability scheduling approach for establishing scheduling policies \cite{aggarwal2017taming, xiang2014joint}.
We use $\omega_{i,j}$ to denote the probability that server $j$ receives the distributed task corresponding to the arrival request of service $i$.
We define the vector $\bm{\omega}_j = \{\omega_{0,j}..., \omega_{i,j}, ...\}, i\in\mathcal{I}$ as the decision for the $j$-th edge server. 
Essentially, we minimize the long-tail effect of distributed edge computation through the decision variable $\bm{\omega}_j, j\in\mathcal{M}$.

Let $B_i$ represent a set of edge servers containing candidate servers that offer service $i$, where $i \in \mathcal{I}$. For instance, user $u_3$ has various parallel computing plans denoted as $B_i = \{(3), (5), (3,5)\}$, to fulfill service requests, provided that both edge servers $\#3$ and $\#5$ are available for $u_3$ as shown in Figure \ref{fig_intro}.
Let $B_i(k)$ denote a set of candidate servers, where the $k$-th element belongs to $B_i$.

Hence, we can deduce $\omega_{i,j}$ from the probability of a parallel plan for implementing service request $i$, expressed as
\begin{equation}\label{formula_Pr_A}
    \omega_{ij} = \sum_{k}  Pr(B_i(k)) \cdot \mathds{1}(j \in B_i(k)),  \forall j \in \mathcal{M}, 
\end{equation}
where $Pr(B_i(k))$ denotes the probability of parallel implementation for service request $i$ by edge servers in $B_i(k)$, and $\mathds{1}(j \in B_i(k))$ is an indicator function that evaluates to $1$ if edge server $j$ is in the set $B_i(k)$ and $0$ otherwise.
%================================================================
% Optimization Problem
%================================================================
\subsection{Problem Formulation}
Let the probability of the occurrence of the tail latency event for all services be $\kappa$.
We denote the probability of tail latency event of the $j$-th edge server by $Pr(T_j \geq \gamma)$, where $\gamma > 0$ represents the tolerance level for tail latency.
% Similarly, we denote $Pr(T_j \geq \gamma)$ as the probability of tail latency event of the $j$-th edge server. 
Applying the complement rule for probability, the occurrence of at least one tail latency event is equivalent to the opposite of no long-tail events occurring, we have that
\begin{align}
    & \kappa =1 - \prod_{j\in\mathcal{M}}(1- Pr(T_j \geq \gamma)). \label{formula_tail_latency_prob}
\end{align}
We formulate the minimization problem of $\kappa$ as follows:
\begin{align}
 &\mathcal{P}:  \quad  \min_{\bm{\omega}_j,\forall j\in\mathcal{M}} \quad \kappa  \\
 & s.t. \quad 0 \leq  Pr(T_j \geq \gamma) \leq 1, 
\end{align}
However, the decision-making variable $\bm{\omega}_j,\forall j\in\mathcal{M}$  lacks a direct connection with the optimization problem. 
The relationship between $\bm{\omega}_j$ and $Pr(T_j \geq \gamma)$ is unknown, making it challenging to solve the optimization problem $\mathcal{P}$.
%================================================================
% Analysis for Tail Latency
%================================================================
\section{Tail Latency Analysis}\label{sec_tail_analysis}
To analyze $Pr(T_j \geq \gamma)$, we employ methods in queueing theory for in-depth analysis $T_j$ and modeling $T_j$ from the \textit{Laplace Transform} technique. 
%================================================================
% Arrival Rate and Average Task Size
%================================================================
\subsection{Service Arrival Rate and Average Size}
Drawing upon queueing theory, we model each service request as a \textit{Poisson Process} \cite{kleinrock1975queueing}. 
Each service request is considered as a computation task, and its average size, denoted as $c_i$, is quantified by the time required for the request to execute on the edge server (e.g., $c_i= 10^7$ CPU cycles/request). 
We assume that when tasks are executed in parallel, the task load is evenly distributed among each edge server.
However, the random variable $c_i$ is closely tied to the computing capability of the edge server (e.g., the parallel task for $u_1$ takes $310$ms on one edge server but only $190$ms on another edge server, as illustrated in Figure \ref{fig_intro}).

Let $\Lambda_j$ represent the service arrival rate collected by the $j$-th server. 
We use $\lambda_i$ to denote the arrival rate of service request $i$, representing the average number of requests arriving per unit time (e.g., $\lambda_i=100$ req/sec) \cite{kleinrock1975queueing}. 
According to the \textit{theory of Poisson splitting} \cite{durrett2019probability}, $\Lambda_j = \sum_i \lambda_i \cdot \omega_{i,j}=\bm{\lambda} \bm{\omega}_j, i\in \mathcal{I}$, where $\bm{\lambda} = {\lambda_0,..., \lambda_i, ...}, i\in\mathcal{I}$.
In Figure \ref{fig_system}, the arrival rate remains consistent across the uplink, server, and downlink queues, denoted as $\Lambda_j$, due to \textit{Burke's theorem} \cite{harchol2013performance, kleinrock1975queueing}.

Therefore, the average size $\bar{c}_i$ of service $i$ at server $j$ is given by 
\begin{equation}
  \bar{c}_i = \frac{\sum_i c_i \lambda_i \omega_{i,j} }{\sum_i \lambda_i \omega_{i,j}}.\label{fomrula_avesize}
\end{equation}
By using matrix
$$C =
(\begin{pmatrix} 
    c_1 &  & \\  
    & ... &  \\
    &  & c_I 
\end{pmatrix}
\bm{\lambda})^\mathrm{T} \bm{\omega}_j
$$ 
to replace denominator of (\ref{fomrula_avesize}), we obtain $\bar{c}_j = \frac{ \bm{\lambda}^\mathrm{T} C \bm{\omega}_j}{\bm{\lambda}\bm{\omega}_j}.$

Let $r_s^j$ denote the server computing capability of the $j$-th edge server, measured as the average number of tasks completed per unit time (e.g., $r_s^j = 10^6$ CPU cycles/ms). 
Similarly, we define the uplink and downlink transmission speeds related to the $j$-th edge server as $r_u^j$ and $r_d^j$. We use $\mu_u^j$, $\mu_s^j$, and $\mu_d^j$ to represent the \textit{average service rates} of the $j$-th uplink, edge server, and downlink.
By the definition of \textit{average service rate} \cite{harchol2013performance}, we employ $r_s^j$ to divide $\bar{c}_j$ to calculate the average number of tasks completed at the $j$-th edge server per unit time (e.g., $\mu_s^j=0.1$ tasks/ms).
Therefore, we have that
\begin{eqnarray}
   \mu^j_u = \frac{r_u^j\bm{\lambda}^\mathrm{T}\bm{\omega}_j}{ \bm{\lambda}^\mathrm{T} C \bm{\omega}_j},  
   \mu^j_s = \frac{r_s^j\bm{\lambda}^\mathrm{T}\bm{\omega}_j}{ \bm{\lambda}^\mathrm{T} C \bm{\omega}_j},  
   \mu^j_d = \frac{r_d^j\bm{\lambda}^\mathrm{T}\bm{\omega}_j}{ \bm{\lambda}^\mathrm{T} C \bm{\omega}_j}.  
\end{eqnarray}
% where $ \Lambda_j = \sum_{i\in K_j} \lambda_i \omega_{i,j} = \bm{\lambda} \bm{\omega}_j$
%================================================================
% Laplace Transform of Response Time
%================================================================
\subsection{Laplace Transform of Response Time}\label{sec_laplace_transform}
By applying the \textit{Laplace Transform} of the $M/M/1$ queue \cite{kleinrock1975queueing}, each response time in the right-hand side of Equation (\ref{formula_response_time}) can be transformed as follows:  
\begin{align}
   \hat{T}_u^j  &= E[e^{-x_j T^j_u}] = \frac{\mu^j_u - \Lambda_j}{\mu_u^j - \Lambda_j +x_j},\\
   \hat{T}_s^j  &= E[e^{-x_j T^j_u}] = \frac{\mu^j_s - \Lambda_j}{\mu_s^j - \Lambda_j +x_j},\\ 
   \hat{T}_d^j  &= E[e^{-x_j T^j_u}] = \frac{\mu^j_d - \Lambda_j}{\mu_d^j - \Lambda_j +x_j}, \label{formula_laplace_transform}
\end{align}
where $x_j$ is the function parameter of the \textit{Laplace Transform}, $\hat{T}_u^j, \hat{T}_s^j$ and $\hat{T}_d^j$ are \textit{Laplace Transform}s of the response time for $T_u^j$, $T_s^j$ and $T_d^j$ respectively.
The response times $T_u^j, T_s^j$ and $T_d^j$ are independent.
Using the \textit{theory of Linearity of Transforms} \cite{harchol2013performance}, we can derive the expression for the \textit{Laplace Transform} of $T_j$:
\begin{equation}
    \hat{T}_j =  E[e^{-x_j T_j}]  = \hat{T}_u^j \cdot \hat{T}_s^j \cdot \hat{T}_d^j. \label{formula_tandem_laplace_transform}
    %= E[e^{-s(T_u^j + T_s^j + T_d^j)}],
\end{equation}
% The response time of a distributed computing depends on the response time of the slowest queue.
% The tail latency event for edge server $j$ happens is equivalent to the union of a sequence of events $T_j > \gamma, j\in\mathcal{M}_i$.  
% We now derive $Pr(L_i\geq \gamma)$ according to $Pr(T_j \geq \gamma), j\in\mathcal{M}$ and Equation (\ref{formula_Pr_A}), obtaining
% \begin{align} \label{formula_latnecy_Li}
%    Pr(L_i \geq \gamma)  & = \sum_j\sum_k Pr(A_i(k)) \cdot \mathds{1}(j \in A_i(k))\cdot Pr(T_j\geq\gamma)  \notag\\
%                         & = \sum_j \omega_{i,j} \cdot Pr(T_j \geq \gamma)
% \end{align}
%================================================================
% Upper-bound of Tail Probability
%================================================================
\subsection{Upper-Bound of Tail Probability}\label{sec_tail_bound}
Suppose we are seeking the optimal solution to problem $\mathcal{P}$ with a given $\bm{\omega}_j, j\in\mathcal{M}$ but an unknown variable $x_j,j\in\mathcal{M}$. 
By leveraging the \textit{Chernoff bound} \cite{mitzenmacher2017probability}, we can formulate the probability bound of $T_j$, given a well-chosen value $x_j \geq 0$,
\begin{equation}
   Pr(T_j > \gamma) \leq \eta_j(x_j),  \label{formula_chenoff_bound}
\end{equation}
where 
\begin{equation}
    \eta(x_j) = \frac{E[e^{x_j T_j}]}{e^{x_j \gamma}}.\label{formula_eta_j}
\end{equation}
% We can obtain the probability bound of long-tail latency for each service by substituting Equation (\ref{formula_eta_j}) into Equation (\ref{formula_latnecy_Li}), and thus
% \begin{equation} \label{formula_service_tail_bound}
%      Pr(L_i \geq \gamma)  \leq \varrho_i,
% \end{equation}
% where
% \begin{equation}
%     \varrho_i(\bm{x})  = \sum_j \omega_{i,j} \cdot \eta(x_j), \bm{x} = \{x_1, x_2, ...,x_M\} 
% \end{equation}

% The moment generating function of $T_j$ is 
% \begin{equation}
%     M_{T_j}(t) = E[e^{tT_j}], t \geq 0.
% \end{equation}
% To state the convexity of $\eta_j(\bm{x})$, we define the $j$-th element in $\bm{x}$ as a vector $\bm{x}_j = \{0,\cdots, \underbrace{x_j}_{j}, \cdots, 0\}$.
% Based on formulas (\ref{formula_tandem_laplace_transform}),(\ref{formula_latnecy_Li)}) and (\ref{formula_chenoff_bound}), we can use $s_j=-x_j, j\in\mathcal{M}$ to substitute $s_j$ in (\ref{formula_tandem_laplace_transform}), and rewrite $Pr(L_j > \gamma)$ and $ \eta(x_j)$ as follows:
% \begin{equation}
%     Pr(L_i \geq \gamma) \leq \varrho_i , 
%     % \frac{1-\rho_j}{1-\rho_j + x}, 
%     \label{formula_latency_bound}
% \end{equation}
% where $\varrho_i = \sum_j \omega_{i,j} \eta_j(x_j), \eta_j(x_j) = \hat{T}_j e^{-x_j \gamma}$,  $x_j$ is the $j$-th element of $\bm{x}$.
Substituting Equation (\ref{formula_tandem_laplace_transform}) into the numerator of Equation (\ref{formula_eta_j}) and considering the first and second derivatives of $\eta_j(x_j)$ with respect to $x_j$ under a given $\bm{\omega}_j$, we have that 
\begin{align}
\eta_j(x_j)          &= \hat{T}_j e^{-x_j \gamma}, \label{formula_etaj} \\
\nabla \eta_j(x_j)   &=  e^{-x_j\gamma} (\nabla\hat{T}_j -\gamma\hat{T}_j), \label{formula_1st_etaj}\\
\nabla^2 \eta_j(x_j) &=  e^{-x_j \gamma} (\gamma^2 \hat{T}_j - 2\gamma \nabla\hat{T}_j + \nabla^2\hat{T}_j)\label{formula_2nd_etaj}.
\end{align}
Moreover, considering $\hat{T}_j$ as a function of $x_j$, we obtain:
\begin{align}
 \hat{T}_j(x_j) &=\frac{\phi_u \phi_s \phi_d}{(\phi_u - x_j)(\phi_s - x_j)(\phi_d - x_j)},\label{formula_Tj} \\
 \nabla \hat{T}_j(x_j) &=\hat{T}_j(x_j) \big(\frac{1}{\phi_u - x_j} + \frac{1}{\phi_s - x_j} + \frac{1}{\phi_d - x_j} \big), \label{formula_1st_Tj} \\
 \nabla^2 \hat{T}_j(x_j) &=  \hat{T}_j(x_j) \big[
        (\frac{1}{\phi_u - x_j} + \frac{1}{\phi_s - x_j})^2 + (\frac{1}{\phi_u - x_j} \notag \\
     &+ \frac{1}{\phi_s - x_j})^2
      + (\frac{1}{\phi_u - x_j} + \frac{1}{\phi_s - x_j})^2
     \big] \label{formula_2nd_Tj}
\end{align}
by replacing $\mu_u^j - \Lambda_j, \mu_s^j - \Lambda_j$ and $\mu_d^j - \Lambda_j$ with $\phi_u, \phi_s$ and $\phi_d$, respectively.
\subsection{Theoretical Analysis}
Given $\phi_u, \phi_s$ and $\phi_d$, we can partition the domain space $x_j > 0$ and analyze the convexity of the piecewise function $\eta_j(x_j)$ on each interval of its domain segments, i.e., $(0, \phi_u), (\phi_u, \phi_s)$ and $ (\phi_s, \phi_d)$. 
\begin{assumption}
$0< \phi_u < \phi_s < \phi_d$. 
\end{assumption}
Note that the following analysis can apply to all the numerical orders of $\phi_u, \phi_s$ and $\phi_d$.
Thus, we assume $0< \phi_u < \phi_s < \phi_d$ to illustrate the following analysis for convenience.

\begin{lemma} \label{lemma_convex_domain}
   $\eta_j(x_j)$ is convex when $x_j \in(0, \phi_u) \cup (\phi_s,  \phi_d)$.
\end{lemma}
\begin{proof}
After substituting the expressions for $\hat{T}_j$, $\nabla \hat{T}_j$, and $\nabla^2 \hat{T}_j$ and rearranging, we obtain:
\begin{align}
  \nabla^2 &\eta_j(x_j) = \frac{\omega_{i,j} e^{-x_j \gamma}\hat{T}_j}{\phi_u\phi_s\phi_d} \cdot Y,
\end{align}
where $Y = \gamma^2 -2\gamma(y_u + y_s + y_d) + (y_u+y_s)^2 +(y_s+y_d)^2 + (y_u+y_d)^2$ by replacing $\frac{1}{(\phi_u - x_j)}, \frac{1}{(\phi_s - x_j)}$ and $\frac{1}{(\phi_d - x_j)}$ by $y_u, y_s$ and $y_d$, respectively. 
However, considering $\gamma$ as the variable, we find that the discriminant of quadratic polynomial $Y$ is $-4y_u^2 -4y_s^2 - 4y_d^2 < 0$ when $x_j \neq \phi_u,  x_j \neq \phi_s,  x_j \neq \phi_d$.
Hence, we can derive $ \frac{\omega_{i,j} e^{-x_j \gamma}}{\phi_u\phi_s\phi_d} \cdot Y > 0$ due to $\omega_{i,j} e^{-x_j \gamma}\hat{T}_j > 0$.
Whether $\nabla^2 \eta_j(x_i)$ is positive or not only depends on the value of $\delta = (\phi_u - x_j)(\phi_s - x_j)(\phi_d - x_j)$.
% We can classify the convex intervals of $\eta_j({\bm{x}_j})$ in the following three situations:
$x_i > 0$ is partitioned as four regions as $(0, \phi_u), (\phi_u, \phi_s), (\phi_s,  \phi_d)$ and $(\phi_d, \infty)$.
If $x_j\in(0, \phi_u)$, $\phi_u - x_j > 0, \phi_s - x_j > 0$, and $\phi_d - x_j> 0$, hence $\delta$ is non-negative, then $\nabla^2 \eta_j(x_i) > 0$.  
If $x_j \in(\phi_s,  \phi_d)$, $\phi_u - x_j < 0, \phi_s - x_j < 0$, and $\phi_d - x_j> 0$, hence $\delta$ is non-negative, then $\nabla^2 \eta_j(x_i) > 0$. 
Thus, $\eta_j(x_i)$ is convex when $x_i \in (0, \phi_u) \cup (\phi_s,  \phi_d)$. 
\end{proof}
% Due to the queueing assumption that service rate should be greater than arrival rate[], $\phi_u > 0, \phi_s > 0$ and $\phi_d > 0$.
% Note that \textcolor{red}{$x_j>0$}$x_j < 0$ such that we can obtain $(\phi_u - x_j), (\phi_s - x_j)$ and $(\phi_d - x_j)$ are non-negative, furthermore $\omega_{i,j} \geq 0$, $\gamma$ is a positive value and $e^{x_j\gamma} > 0$.
% Hence, by using (\ref{formula_1st_order}) and (\ref{formula_2nd_order}), $\nabla^2 \eta_j(\bm{x}_i) > 0$, subsequently $\eta_j(\bm{x}_i)$ is convex, according to the second-order condition lemma of convex function\cite{}. 
% Followed by the fact that non-negative sum of convex functions is convex, we conclude that $\eta_j(\bm{x}) = \sum_j \eta_j(\bm{x}_j)$ is convex.
However, $\hat{T}_u^j, \hat{T}_s^j$ and $\hat{T}_d^j$ are non-negative due to the definition of the Laplace Transform, which is necessarily positive based on (\ref{formula_tandem_laplace_transform}).
Hence, $x_j$ should satisfy $\frac{\phi_u}{\phi_u - x_j} > 0$, $\frac{\phi_s}{\phi_s - x_j}>0$ and $\frac{\phi_d}{\phi_d - x_j}>0$ simultaneously, we have that $x_j\in(0, \phi_u)$ according to $0< \phi_u < \phi_s < \phi_d$.
We conclude that from Lemma \ref{lemma_convex_domain}, the only feasible domain of $x_j$ is convex, i.e., $(0, \phi_u)$.
Therefore, by using the first-order condition of convex function \cite{boyd2004convex}, the minimizer of $\eta_j(x_j)$ is $x_j^\star$ when $\nabla\eta_j(x_j^\star) = 0, x_j^\star\in (0, \phi_u)$.
% We formulate the minimized problem $\mathcal{P}$ for the tail latency of service $i\in\mathcal{K}$ as
% \begin{align} 
%    \mathcal{P}:\quad &\min \quad \eta_j, \\
%    s.t. &\quad 0\leq x_j<\phi_u,  \\
%         &\quad 0< \eta_j <1, \forall j\in\mathcal{M}
% \end{align}

% \begin{theorem}\label{theorem_first_order}
%     $\bm{x}^\star = \{x_1^\star, x_2^\star, ...,x_M^\star\}$ is a global minimizer of $\varrho_i(\bm{x})$ by solving $\nabla\eta_j(x_j^\star) = 0, \forall j\in\mathcal{M}$ with given $\bm{\omega}_j$ and $\gamma$. 
% \end{theorem}
% \begin{proof}
% For given non-negative coefficients as vector $\bm{\omega}_i$, $\varrho_i = \sum_j \omega_{i,j} \eta_j(x_j)$ is convex by applying the operation that non-negative weighted sums of convex functions preserve convexity\cite{boyd2004convex}.
% Hence, the optimal value of the first condition of $\nabla\varrho_i = \sum_j \omega_{i,j} \nabla\eta_j(x_j^\star) =0$ leads to the optimal solution of $\varrho_i$.  
% Therefore, $\bm{x}^\star = \{x_1^\star, x_2^\star, ...,x_M^\star\}$ is the optimal solution of $\varrho_i$.
% \end{proof}
\begin{theorem}\label{theorem_upper_bound}
By applying the rule of \textit{Chernoff bound}, $1 - \prod_{j\in\mathcal{M}}(1-\eta_j(x_j^\star))$ serves as an optimal upper bound for the probability of tail latency $\kappa$ with given $\bm{\omega}_j, j\in\mathcal{M}$ and $\gamma$.
\end{theorem}
\begin{proof}
$1-\eta_j(x_j^\star)$ ensures the maximum probability of non-long-tail latency occurrence due to the minimization of $\eta_j(x_j^\star)$. 
Consequently, the product of $\prod_{j\in\mathcal{M}}(1- \eta_j(x_j^\star))$ denotes the maximum probability of concurrent non-long-tail latency for any service $i$. 
Conversely, when $1 - \prod_{j\in\mathcal{M}}(1- \eta_j(x_j^\star))$ is considered, it represents the minimum probability of at least one service $i$ encountering long-tail delays. 
This value establishes the optimal upper bound for the long-tail probability of distributed edge computation.
\end{proof}
%================================================================
% Tail Modeling
%================================================================
\section{Tail Learning}\label{sec_tail_learning}
Although we have identified the optimal upper bound for $\kappa$ as per Theorem \ref{theorem_upper_bound}, problem $\mathcal{P}$  remains challenging to solve due to the non-convex nature concerning the variable $\bm{\omega}_j, j\in\mathcal{M}$. 
The absence of a readily available solver makes non-convexity problems extremely difficult to tackle.

Therefore, in light of the upper bound analysis for problem $\mathcal{P}$ discussed above, we leverage a reinforcement learning (RL) algorithm to comprehensively exploit hidden metrics associated with tail latency through statistical learning. 
This approach allows us to obtain a sub-optimal solution $\bm{\omega}_j, j\in\mathcal{M}$ and the corresponding $x_j$ until the RL converges.
% We propose an approach to learn tail latency by reinforcement learning.
% However, in a dynamic system, $\omega_{i,j}$ in (\ref{formula_Pr_A}) is brutal to solve due to the uncertainty of $Pr(A_{i,h}^l)$, in which $A_{i,h}^l$ offers the parallelism to support service $i$.
% If we fix $Pr(A_{i,h}^l)$ in the system, the long-tail event will dramatically arise as time goes on than a system with an adaptive $Pr(A_{i,h}^l)$. 
% To achieve the robustness and efficiency of the designed algorithm, we apply a reinforcement learning approach to generate online decisions $\omega_{i,j}$ according to the dynamics of service requests and adaptive $A_{i,h}^l$.
%================================================================
% State
%================================================================
\subsection{Reinforcement Learning-based Method Design}
\textbf{State.}\label{sec_enhanced_state}
% Considering the tandem queueing network in Figure \ref{fig_system}, 
Since the computing capabilities of edge servers are limited, once the arrival rate of task requests exceeds the service rate of edge servers, the queue backlog increases significantly. 
The backlog of tandem queuing tasks will cause a notable latency in response time $T_j$. 
Therefore, we utilize the maximum($q_{max}^j$), minimum($q_{min}^j$), average($q_{ave}^j$), and variance($q_{var}^j$) of the queue length as state variables, where $q_{max}^j, q_{min}^j, q_{ave}^j, q_{var}^j$ are three-dimensional vectors respectively, each dimension representing uplink, edge server and downlink in the $j$-th tandem queueing network.  
Furthermore, employing \textit{feature engineering} techniques \cite{khurana2018feature} in the edge computation system enables us to obtain a nuanced representation of the system state.
In control engineering for nonlinear dynamic systems \cite{ogata2010modern}, differential equations are often employed to model the state space for improved system identification. 
Drawing inspiration from this, we use the response time of each $T_j$ and the upper bound probability of tail-latency $\eta_j(x_j)$, along with their first-order and second-order expressions, as the state for reinforcement learning to comprehensively capture the state characteristics of the distributed edge computation system.
Besides, incorporating the relevant information representing the queues will enhance the comprehensiveness of the system state representation. 
Our proposed algorithm combines $\eta_j(x_j)$, $\nabla\eta_j(x_j)$, $\nabla^2 \eta_j(x_j)$ in (\ref{formula_etaj})-(\ref{formula_2nd_etaj}) and $\hat{T}_j(x_j)$, $\nabla \hat{T}_j(x_j)$, $ \nabla^2\hat{T}_j(x_j)$ in (\ref{formula_1st_Tj})-(\ref{formula_2nd_Tj}) to define the state as $s_j$  for the $j$-th tandem network. 

The observation period is the interval $\Delta$ defined in Section \ref{sec_reward}.
We then denote the state $S$ of reinforcement learning by
\begin{align}
S = &\{..., S_j, ..\} , S_j =\{q_{max}^j, q_{min}^j, q_{ave}^j, q_{var}^j,  \\ \nonumber
&\eta_j(x_j), \nabla\eta_j(x_j), \nabla^2 \eta_j(x_j), \hat{T}_j(x_j), \nabla \hat{T}_j(x_j),  \nabla^2\hat{T}_j(x_j)\}  \\ \nonumber
\end{align}
% \textcolor{blue}{TODO: change to the average queue length}Thus, we use a vector $\mathbf{s}$ with the dimension $3M*K$ to represent each queue length of service $k\in\mathcal{K}$ at each uplink,  server and downlink queue $j\in\mathcal{M}$.
% We also use $\rho^j_u, \rho^j_s, \rho^j_d$ to denote device utilization, which can be calculated as $B/\tau$ where $B$ is the non-idle time during $\tau$, $\tau$ is the observation period.
% Device Throughput $X^j_u, X^j_s, X^j_d$, $F/\tau$, $F$ is the number of tasks completed during $\tau$.
% \textcolor{blue}{TODO:}Average response time $\bar{T}^j_u(k), \bar{T}^j_s(k), \bar{T}^j_d(k)$, max response, min response, variance of response time.
% Average waiting time $\bar{W}^j_u(k), \bar{W}^j_s(k), \bar{W}^j_d(k)$.

% the service rate of each uplink $\{\mu_u^1, \mu_u^2, \cdots, \mu_u^M \}$,
% the service rate of each server $\{\mu_s^1, \mu_s^2, \cdots, \mu_s^M \}$,
% and the service rate of each server $\{\mu_d^1, \mu_d^2, \cdots, \mu_d^M \}$.

%================================================================
% Action
%================================================================
\noindent\textbf{Action.}
% Denote $e_{i,j}$ as an indicator for deploying service $i$ to server $j$.
Our proposed algorithm outputs the probability scheduling decision, as detailed in Section \ref{sec_decision_Ai}.
We define the action vector of reinforcement learning to be $A = \{..., \bm{\omega}_j\, ...\}, j\in\mathcal{M}$.

%================================================================
% Reward
%================================================================
\noindent\textbf{Reward.}\label{sec_reward} 
Each step is defined by $[t, t+\Delta]$.
We denote $d_n$ as the latency of service requests $n$.
Let $\Xi_t$ denote as the set of arrivals of user requests from time step $t$, i.e., $[t, t + \Delta]$, excluding arrivals during $[t+\Delta-\gamma, t+\Delta]$, which do not leave until $t+\Delta$, where $\Delta \gg \gamma$ \cite{raeis2021queue}.
We assign the reward to each service request latency based on the comparison to $\gamma$ as:
\begin{eqnarray}
    \phi_n = 
    \left\{
        \begin{aligned}
         & \beta_1, d_n < \gamma, \\
         & \beta_2, d_n \geq \gamma, 
        \end{aligned}
    \right. 
\end{eqnarray}
where $\beta_1$ and $\beta_2$ are two positive parameters, $n \in \Xi_t$ is the index of a request arrival.
Specifically, service requests arriving after time $t$ are often queued within $[t,t+\Delta]$. 
In our proposed algorithm, we need to consider the long-tail delay due to task backlog, so we use the task queue length as a penalty term.
To mitigate congestion within the system queue, we address the growth of queue lengths for the uplink, server, and downlink during two consecutive time intervals, denoted as $\Delta$. We introduce penalties for these queue length increases, expressed as $\beta_3 \epsilon_u^j, \beta_3 \epsilon_s^j$, and $\beta_3 \epsilon_d^j$ for $j \in \mathcal{M}$, where $\beta_3$ is a positive parameter.
Hence, we denote the reward function by 
\begin{equation} \label{formula_reward}
R = \sum_{n\in\Xi_t}  r_n,  
\end{equation}
where
\begin{eqnarray}
  r_n = \beta_1 \mathds{1}\{d_n < \gamma \}- \beta_2 \mathds{1}\{d_n \geq \gamma\} - \beta_3(\epsilon^j_u + \epsilon^j_s +   \epsilon^j_d).
\end{eqnarray}
%================================================================
% Agent Framework
%================================================================
\subsection{Autonomous Edge System}
\begin{figure}[htbp]
    \centering
    \includegraphics[width=0.8\textwidth]{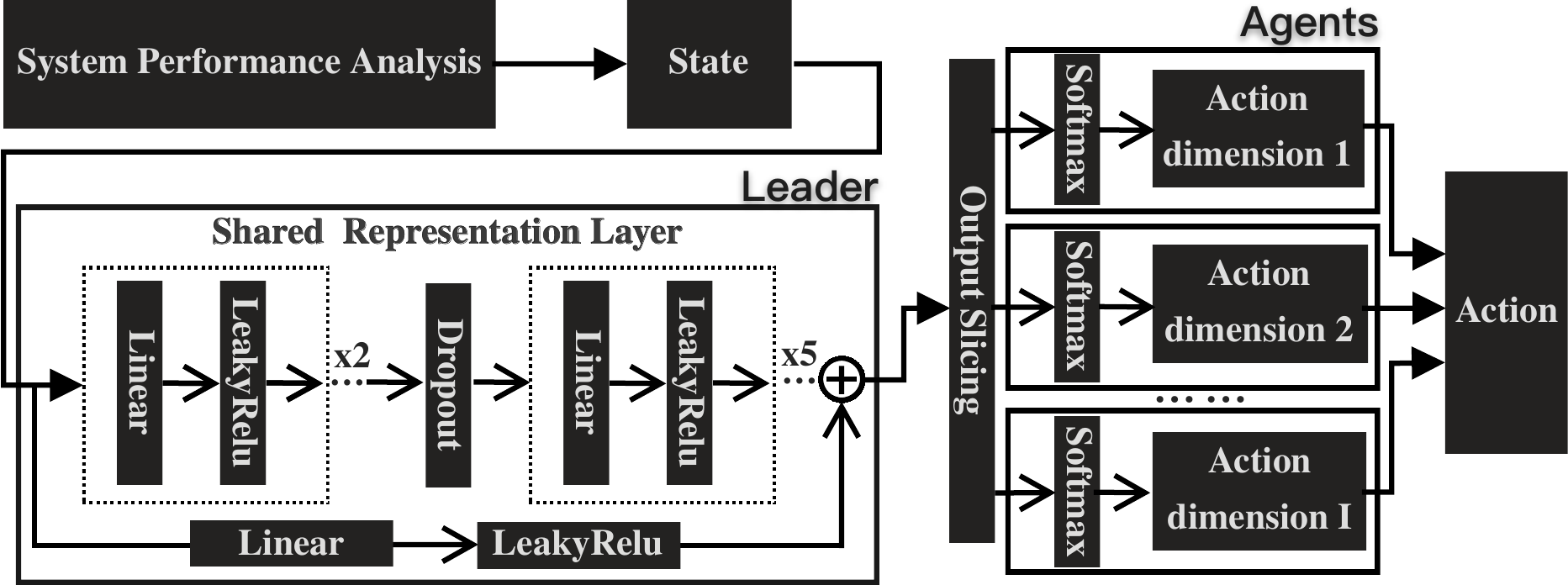} 	
    \caption{Autonomous Agents framework for learning procedure.}
    \label{fig_algo_framework}
\end{figure}

After defining states, actions, and rewards above, we describe the overall framework of our proposed algorithm in Figure \ref{fig_algo_framework}.
We can initialize the \textit{autonomous edge system} by utilizing synchronous network \textit{election algorithms}, such as the \textit{LCR} algorithm for edge servers in a ring connection and the \textit{FloodMax} algorithm for any connection topology \cite{bullo2009distributed}.
Consequently, a single edge server \textit{leader} in the edge computing network hosts the \textit{shared representation layer}, while a subset of others deploys parallel service decision-making \textit{agents} as shown in Figure \ref{fig_algo_framework} \cite{nguyen2023learning}.
In summary, after the completion of the \textit{leader} node and each \textit{agent} node election, the network is initialized, and each \textit{agent} calculates the decision for its corresponding service based on the shared output of the \textit{leader} calculation node.
In our system, it is possible to select the leader server and each agent server even without using the election algorithm. 
We can manually preset the edge servers corresponding to these two roles in the edge network. 
After completing the initial setting mentioned above, as illustrated in Figure \ref{fig_algo_framework}, the \textit{system performance analysis} module can provide the enhanced state as the input to the \textit{shared representation layer} of the leader node. 
The calculated results from the \textit{shared representation layer} are transmitted to each agent node, where each \textit{agent} node computes the action decision corresponding to the service.
We present the detailed adaptive decision-making process as follows.
% This paper assumes that the leader and agent nodes are already available and focuses on the design of the decision-making algorithm.
The module labeled \textit{System Performance Analysis} in Figure \ref{fig_algo_framework} produces an enhanced state that unveils performance metrics in depth. 
The calculation process of this module primarily involves the following:
\begin{itemize}
    \item 
    The introduction of response time from the tandem queuing network in Section \ref{sec_res_t}.
    \item 
    The probability upper bound of the tail latency from Section \ref{sec_laplace_transform}.
    \item 
    Their first-order and second-order mathematical expressions (Equations (\ref{formula_etaj}) to (\ref{formula_2nd_Tj})).
\end{itemize}

From the definition of parallel computing plans in Section \ref{sec_decision_Ai}, it is evident that each service request can utilize various combinations of edge servers where the service is deployed for task execution. 
However, with an increasing variety of services, the action space faces a combinatorial explosion. 
The system's action space dimension will reach $|B_i|^I$ under the condition of one-hot coding. 
One approach is to model a distributed edge computation system that maintains multiple reinforcement learning agents to perform decision-making learning on parallel computing plans for each service. 
However, this leads to a lack of collaboration between the decision spaces corresponding to different actions for services because the coupling relationship between them in a global learning process for all actions needs to be considered.
To address this challenge, we introduce an autonomous edge computation system comprising a shared structure of reinforcement learning and multiple distributed service decision-making \textit{agents}. 
The system appoints a leader, determined through \textit{election algorithms} across all edge servers, responsible for the \textit{Shared Representation Layer} in Figure \ref{fig_algo_framework}, inspired by the design of the \textit{Action Branching Architecture} \cite{tavakoli2018action}.  
% To address this challenge, we introduce a \textit{Shared Representation Layer} in Figure \ref{fig_algo_framework}, inspired by the design of the \textit{Action Branching Architecture}\cite{tavakoli2018action}. 
This module utilizes the enhanced state as input to generate collaborative actions for all decisions related to service requests.
Specifically, our experimental results suggest that incorporating the dropout module, along with a module similar to \textit{ResNet} \cite{langford2021enki}, in the layers of the \textit{Shared Representation Layer} provides advantages for the learning performance of the model.
This placement helps prevent the model from encountering issues of early vanishing gradients \cite{wu2015towards}.

Besides, each specific service designates an \textit{agent}. 
This agent receives output from the leader to autonomously determine the execution of the specific service requests on the relevant edge servers. 
During algorithm runtime, the \textit{Shared Representation Layer}'s output is conveyed to each service action branch via the \textit{Output Slicing} module, facilitating the selection of the corresponding action from $B_i$.
Receiving input from \textit{Softmax} in each agent $i \in \mathcal{I}$, \textit{Action dimension $i$} module outputs the mapping of the probability distribution $Pr(B_i(k))$ of each service's parallel plan $B_i(k)$ (Section \ref{sec_decision_Ai}) to the decision $\omega_{i,j}$ for the $i$-th service and $j$-th edge server. 
This mapping is precisely calculated through Equation (\ref{formula_Pr_A}). 
Ultimately, \textit{agent} $i$ produces action $\omega_{i,j}, j\in\mathcal{M}$, which constitutes action $A$ with a space dimension \(\sum_{i=0}^I{|B_i|}\).
%================================================================
% Tail-learning Algorithm
%================================================================
\subsection{Tail-learning Algorithm}
Let $\theta$ be the parameters of the neural networks used for training reinforcement learning. 
We denote the performance objective for reinforcement learning as $J(\theta) = \mathbb{E}[R | \pi]$, which is the average expectation of the total reward under policy $\pi$.
% Our proposed tail-learning problem should satisfy the chance-constrained formula (\ref{formula_all_services_bound}).
% Therefore, to maximize $J(\theta)$, we need to solve the optimization problem $\mathcal{P}$ as follows:
% \begin{eqnarray}
%     &\max_{\theta}&  J(\theta), \\
%     &s.t.& G(\theta) \leq \kappa, 
% \end{eqnarray}
% where $G(\theta) =  1 - \prod_{i\in\mathcal{K}} (1 - Pr(L_i \geq \gamma))$.
% Moreover, the Lagrangian function of $\mathcal{P}$ is 
% \begin{eqnarray}
%      L(\theta, \xi)& = J(\theta) + \xi(G(\theta) - \kappa),
% \end{eqnarray}
% where $\xi$ is a Lagrange multiplier. 
% Problem $\mathcal{P}$ is equivalent to the saddle-point optimization\cite{} given by
% \begin{equation}
%    \min_{\xi}\max_{\theta} L(\theta, \xi) 
% \end{equation}

% However, due to the stochastic task arrivals, we can not directly solve the problem by the primal-dual gradient descent\cite{}. 
% Thus, we apply a risk-sensitive reinforcement learning method\cite{} to handle the measurement of the probability of tail events. 
% We propose a two stages \textit{Tail-learning} method in Algorithm \ref{algo_rl} by descending in $\xi$ and ascending in $\theta$ using $\nabla_{\theta} L(\theta, \xi)$ and $\nabla_{\xi} L(\theta, \xi)$. 
% To estimate $\nabla_{\theta} L(\theta, \xi)$ and $\nabla_{\xi} L(\theta, \xi)$, let $\hat{\nabla}J(\theta)$ be the estimator of $\nabla J(\theta)$,  $\hat{\nabla}G(\theta)$ be the estimator of $\nabla G(\theta)$, and $\hat{G}(\theta)$ be the estimator of $G(\theta)$, respectively.
Algorithm \ref{algo:rl} outlines the procedures of \textit{Tail-learning} method based on \textit{REINFORCE} \cite{williams1992simple}.  
Our proposed algorithm runs $K$ episodes and $N$ steps for each episode.
According to the principle of the \textit{Markov chain} \cite{durrett2019probability}, the distributed edge computation system will renew the state, action, and corresponding reward in each step.
We use \textit{Monte Carlo Policy Gradient} \cite{sutton2018reinforcement} to update the parameter in the neural network during each episode.
While fixing the policy parameter $\theta$ in a sample trajectory, line $\#\ref{algo_state}$ gives a new state by using Equations (\ref{formula_etaj})-(\ref{formula_2nd_Tj}) . 
Line $\#\ref{algo_action}$ derives the action, which represent the decision for each service $i\in\mathcal{I}$ to allocate to each server $j\in\mathcal{M}$.
When all steps are completed, line $\#\ref{algo_trajectory}$ will collect a sequence containing state, action, and reward as a trajectory of reinforcement learning.
Using the trajectory, line $\#\ref{algo_reward}$ receives the reward of each step and then accumulates it with a discount factor before the next episode.
Due to the designed agent framework based on \textit{Action Branching Architecture}, each \textit{Action dimension \(i\)} module calculates its log probabilities as \(\nabla\log \pi(A_n^i | S_n, \theta), \forall i\in\mathcal{I}\). 
Finally, we aggregate them following line \(\# \text{ \ref{algo_update_parameter}}\) to update parameter \(\theta\) by the gradient ascent principle.

\RestyleAlgo{ruled}
%% This is needed if you want to add comments in
%% your algorithm with \Comment
\SetKwComment{Comment}{/* }{ */}
\begin{algorithm}[hbt!]
\caption{Tail-learning}
\label{algo:rl}
\KwData{
Step size: $\alpha$, Discounted factor: $\sigma$, Initial neural network parameter: $\theta$, 
$d_n, \forall n \in \Xi_t, t = i * \Delta , i \in \mathbb{N}_{+} $}
\For{$k \gets 1 $ to $K$}{ 
    \Comment{According to (\ref{formula_etaj})-(\ref{formula_2nd_Tj}), we can derive the state.}
    \Comment{Using on $S_n$ as input,we can obtain the action.}
    \Comment{Within $[t, t+ \Delta]$, we can add up the reward according to $(\ref{formula_reward})$.}
    \For{$n \gets$ $0$ to $N$}{
    Observe state $S_n = \{...,\eta_j(x_j), ..., \nabla\eta_j(x_j)..., $         \label{algo_state}
    $ ...,\nabla^2 \eta_j(x_j), ..., \hat{T}_j(x_j), ...\nabla \hat{T}_j(x_j)..., \nabla^2\hat{T}_j(x_j), $
    $...,q_{max}^j, q_{min}^j, q_{ave}^j, q_{var}^j,... \}, j\in\mathcal{M}$\\ 
    Calculate action $A_n$. \\ \label{algo_action}
    Receive reward $R_n$.  \label{algo_reward}
    }
Collect trajectory $S_0, A_0, R_1, ..., S_{n-1}, A_{n-1}, R_n ,...,   S_{N-1}, A_{N-1}, R_N$.\\ \label{algo_trajectory}
\For{$n \gets$ $0$ to $N-1$}{
Discounted reward per step: $G = \sum_{m = n+1}^{N} \sigma^{m-n-1} R_m$\\  \label{algo_disc_reward}
Update: $\theta = \theta + \alpha \sigma^n G \nabla \sum_{i\in\mathcal{I}}\log \pi(A_n^i | S_n, \theta)$ \label{algo_update_parameter}
}
}
\KwResult{Policy $\pi(\theta | s)$}
\end{algorithm}
% $|\Xi_{t}|$ is the cardinality of $\Xi_{t}$.
%================================================================
% Evaluation
%================================================================
\section{Evaluation} \label{sec_evaluation}
%================================================================
% Settings
%================================================================
\subsection{Tail-learning Settings}
This section outlines the specific settings employed to evaluate our proposed algorithm. 
We conduct experiments using an \textit{AMD EPYC 7T83 CPU@2.4GHz} and an \textit{NVIDIA GeForce RTX 4090} graphical card.
The parameters were carefully chosen, and the network architecture was designed to capture the complexities of the distributed edge computing environment.
We conducted our evaluations over $4,000$ episodes, with each episode comprising $N=15$ steps. 
Based on experimental results and experience,
we set the reward parameters as \( \beta_1 = 0.1\), \( \beta_2 = 0.3\), and \( \beta_3 = 0.1\), the discounted factor \(\sigma\) for our experiments to \(0.99\), 
the layer size of the linear module in our algorithm  to \(2048\), 
the dropout probability parameter to $0.2$,
the threshold for tail latency \(\gamma\) to \(40\). 
We utilized the \textit{squared L2 normalization} loss module during the training process \cite{bengio2017deep}. 
We employ the \textit{Adam} optimizer for model optimization and set the learning rate for our experiments to \(1 \times 10^{-4}\). 
% These settings are standardized across our evaluations to ensure consistency and enable reproducibility of results. 
%================================================================
% Simulator Settings
%================================================================
\subsection{Simulator Design}
To thoroughly assess our algorithm, we have developed a high-performance, high-precision modular emulator.
This section briefly outlines the design and implementation of our simulator. 
% When streaming video files over the Internet, it is often necessary to convert the original video file into a specific format by altering bitrate, encoding algorithms, file containers, etc., to accommodate varying Internet connections and end-user device environments. 
% Our focus is specifically on \textit{transcoding} M2TS Blu-ray videos into the Matroska format \cite{noe2007matroska}.
We utilize numerical simulation methods to generate the dataset, incorporating eight services where service requests follow a Poisson distribution determined by the average arrival rate and task size specified in Table \ref{tab_arr_rate_ave_size}.
% designed to emulate FFmpeg transcoding \cite{tomar2006converting}. 
% the source code of which is publicly accessible at \url{https://github.com/dynos01/server_simulator}.
Therefore, we construct the simulator for the \textit{distributed video transcoding} scenario based on the above design, implementing functional components using \textit{Rust}\footnote{https://www.rust-lang.org/}. 
The primary components include a global coordinator, edge servers, and a reinforcement learning interface.
\begin{itemize}
    \item 
    \textit{Global Coordinator: }Operated by a high-precision global clock, the global coordinator oversees task allocation to edge servers at intervals defined in data source files. 
    It also manages the synchronization of sub-tasks and the collection of metrics. 
    \item
    \textit{Edge Servers: }The edge servers, functioning as threads, perform the actual computations and generate metrics for the coordinator to gather and analyze. 
    \item
    \textit{Reinforcement learning interface: }The reinforcement learning interface, a collection of HTTP-based endpoints, integrates with our proposed tail-learning to provide near real-time server environment data and implement actions generated by the algorithm.
\end{itemize}
% The emulator and its associated tools are implemented in approximately $2,000$ lines of \textit{Rust} code. 
%================================================================
% Edge Severs and Services
%================================================================
\subsubsection{Edge Environment Parameters}
\begin{table}[h!]
\centering
\begin{tabular}{||c c c ||} 
 \hline\hline
  Edge server          &$r_u^j, r_s^j, r_d^j$ ($10^6$ cycles/ms)   &Supported services(type index)     \\
  $1$-th               &$[5.4, 7.2, 5.4]$                          &$\#1,\#2,\#3,\#8$                      \\  
  $2$-th               &$[7.0, 8.0, 6.0]$                          &$\#1,\#5,\#6,\#8$                       \\   
  $3$-th               &$[8.0, 8.7, 8.0]$                          &$\#1,\#3,\#6,\#8$                       \\
  $4$-th               &$[5.3, 6.5, 4.5]$                          &$\#2,\#4,\#5,\#7,\#8$                       \\[1ex]
 \hline
\end{tabular}
\caption{Computing capability v.s supported services at edge servers.}
\label{tab_computng_capability_supported_services}
\end{table}
In the numerical experiment, we configure four edge servers with varying computing capabilities, as detailed in the second column of the Table \ref{tab_computng_capability_supported_services}. 
Each edge server is capable of supporting different types of services as illustrated in the third column.
%================================================================
% Simulation Data Set
%================================================================
\subsubsection{Simulation Data Set}
\begin{table}[h!]
\centering
\begin{tabular}{||c c c c c c c c c c ||} 
 \hline\hline
 services types      &  \#1 & \#2  & \#3& \#4& \#5& \#6& \#7& \#8 &\#9  \\
 $\lambda_i$($100$ request/sec)       &  4 & 5  & 6& 3& 4& 5& 4& 4  &3 \\
 $c_i$($10^7$cycles/request)       &  4.1 &4.0  & 4.2& 4.9& 4.0& 4.0& 4.2& 4.5 &4.9  
 \\ [1ex] 
 \hline
\end{tabular}
\caption{Service arrival rate v.s average task size.}
\label{tab_arr_rate_ave_size}
\end{table}
We note that the maximum value of the response time in Figure \ref{fig_motivation} is nearly $50$. 
Therefore, we employ this maximum value as the workload required for the task with the largest size in the simulation data. 
Using 50 as the maximum value, we randomly sample the average CPU cycle value $c_i, i\in\mathcal{I}$ required for nine services based on $U(40, 50)$. 
Then, we amplify them by a reasonable order of magnitude while retaining two significant digits. 
Considering that the distributed video transcoding task in Figure \ref{fig_motivation} has an average approximate rate of $460$ requests/sec , we assume that the average service rate for each of the nine services $\lambda_i,\in\mathcal{I}$ is randomly sampled from $U(300, 620)$, retaining one significant digit.
Thus, we utilize numerical simulation methods to generate the dataset, incorporating nine services where service requests follow a Poisson distribution determined by the average arrival rate and task size specified in Table \ref{tab_arr_rate_ave_size}.
Moreover, each service in the dataset is randomly composed of $200,000$ to $400,000$ service requests.
%================================================================
% Simulator Designing
%================================================================
% \subsubsection{Simulator Design}
% We develop a simulation system using Rust, implementing real-time experiments by simulating computing tasks, edge servers, uplink transmission queues, and downlink transmission queues. 
% The simulation is conducted based on the parameter settings of the parallel computing system outlined in Table \ref{tab_arr_rate_ave_size} and Table \ref{tab_computng_capability_supported_services}. 
% The time unit during simulation is $0.02$ms.
\subsection{Benchmarking Policies}
\begin{enumerate}
    \item \textit{Randomized algorithm (RA)}:
    The randomized algorithm employs the principle of random selection for a tuple in \(B_i\) of the \(i\)-th service request, thereby allocating a parallel computing plan represented by the tuple for the service request.
    \item \textit{Greedy algorithm (GD)}:
    The greedy algorithm selects the tuple within \(B_i\) of the \(i\)-th service request, containing the maximum number of edge servers, thereby allocating a greedy parallel computing plan represented by the tuple for the service request.
    \item \textit{Delay-aware selection algorithm (DA) \cite{xie2018cutting}}:
    The delay-aware selection scheme aims to mitigate tail latency by assigning service requests to the faster edge servers by utilizing the tandem queuing network's queue length to predict each edge server's response delays.
    \item \textit{Queue-Learning Approach (QLA) \cite{raeis2021queue}}:
    QLA leverages reinforcement learning with queue length as the state to enhance QoS in a tandem queueing network.
    % \item Tail-bound based approximation
    % \item Queue based reinforcement learning 
\end{enumerate}
%================================================================
% Performance
%================================================================
\subsection{Performance}
%================================================================
% CDF for Tail Tatency
%================================================================
\subsubsection{CDF of Tail Latency}
\begin{figure}[htbp]
        \includegraphics[width=0.32\linewidth]{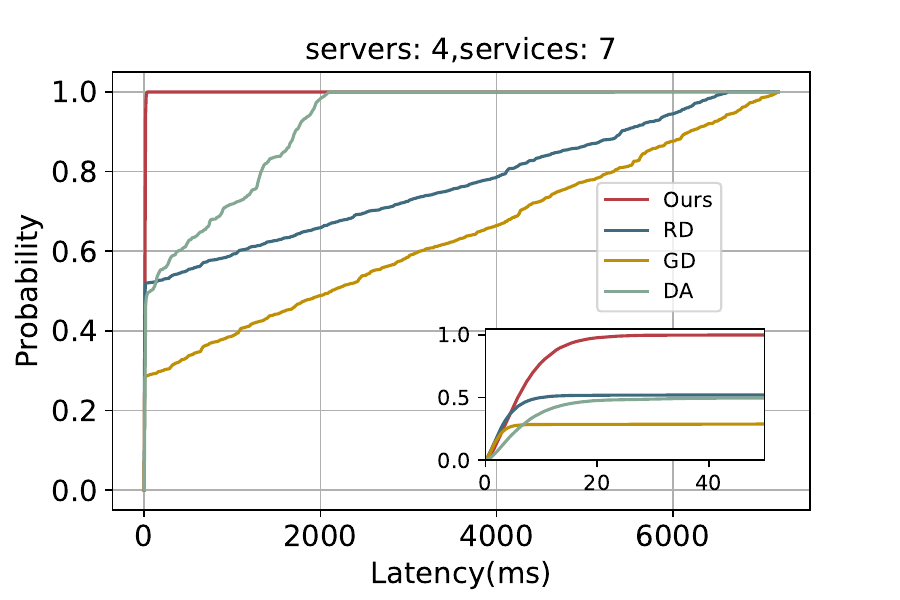}
        \includegraphics[width=0.32\linewidth]{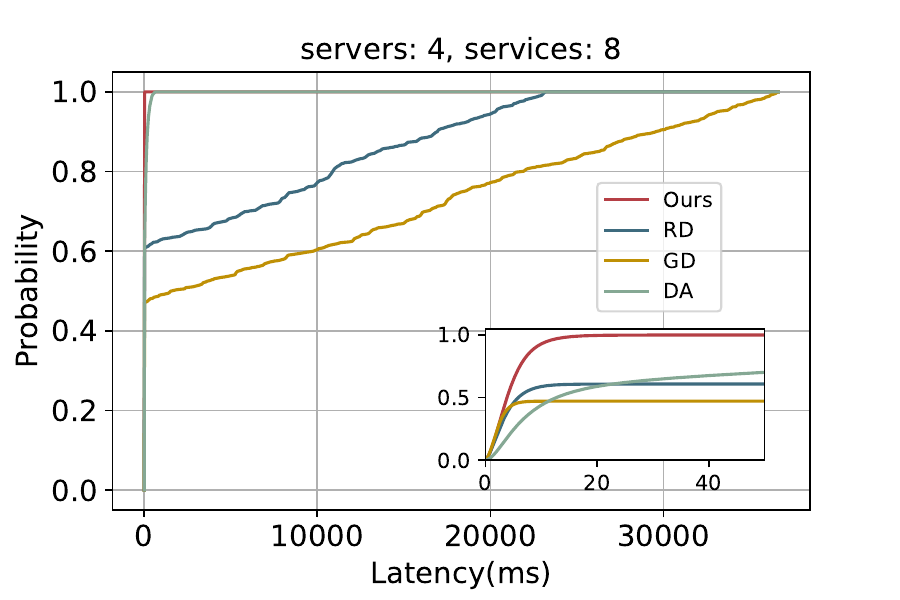}
        \includegraphics[width=0.32\linewidth]{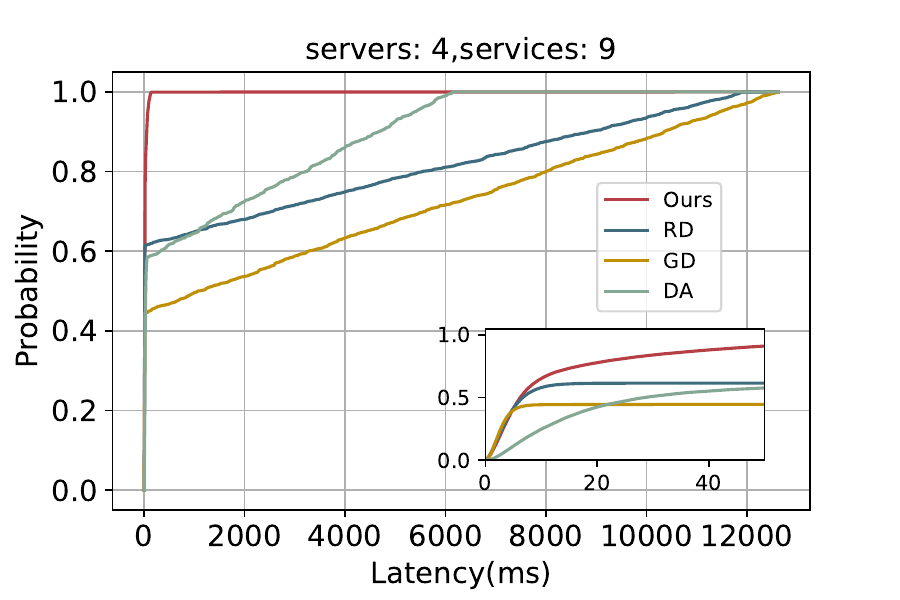}
        
        \includegraphics[width=0.32\linewidth]{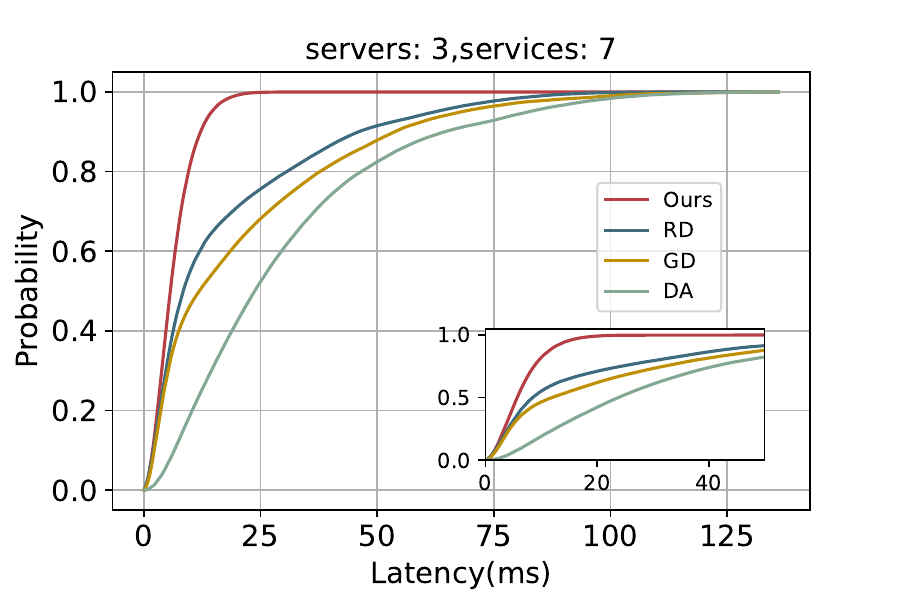}
        \includegraphics[width=0.32\linewidth]{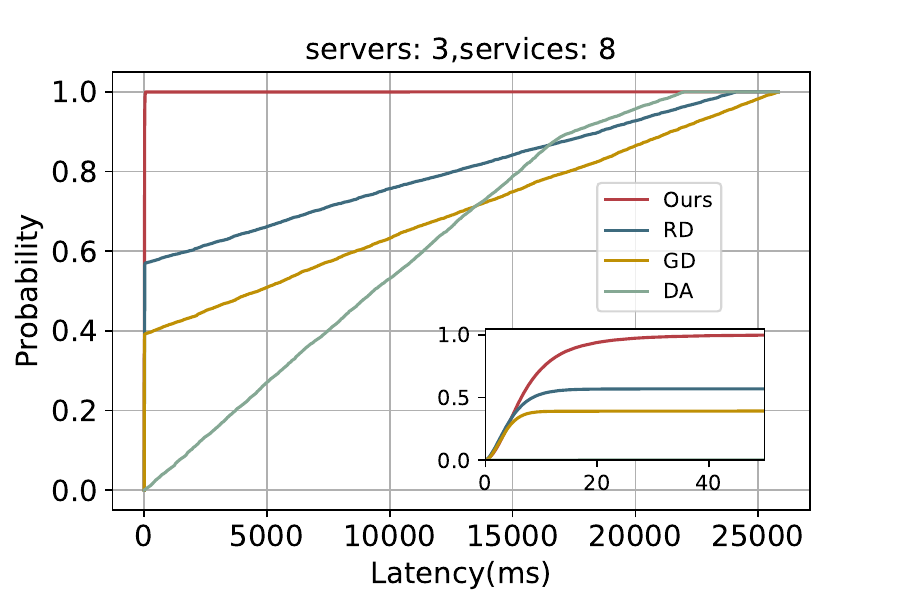}
        \includegraphics[width=0.32\linewidth]{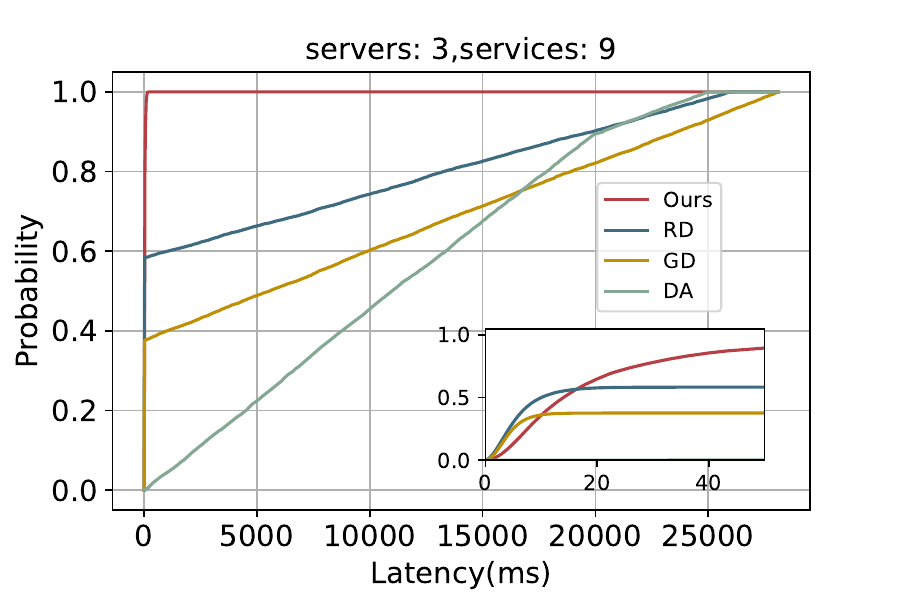}
    \caption{\textit{CDF}s for configurations with 3 and 4 servers under various service type combinations.}
    \label{fig:tail_latency}
\end{figure}
In the performance evaluation of a distributed edge computing environment with a fixed number of servers and services, we utilized three and four edge servers with seven, eight, and nine different service combinations to simulate the response time of distributed parallel service requests. 
Figure \ref{fig:tail_latency} displays the response time as a \textit{Cumulative Distribution Function (CDF)} diagram \cite{durrett2019probability}.

It can be observed from the partial enlargement of each figure that the \textit{CDF} of the response time of our proposed algorithm converges to probability one faster than all comparison algorithms, regardless of the network environment.
In the Top row of Figure \ref{fig:tail_latency}, moving from left to right, the algorithm performance comparison subfigures illustrate the increase in service types from $7$ to $9$ within the edge network of four servers. 
Our proposed algorithm consistently performs optimally as various services expand in the same edge distributed computation system. 
Notably, the \textit{CDF} value of our algorithm tends to reach $1$ at an earlier response time, indicating a faster convergence than other algorithms.
Moreover, considering the random selection of service types supported by servers in three edge environments, the first two subfigures in the top row suggest that adding a service type does not necessarily exacerbate the tail latency. 
This is evident from the \textit{CDF} of the task completion time under our proposed algorithm in the second subfigure, which approaches $1$ more rapidly than the first subfigure.
But, the latter two subfigures in the top row demonstrate that even with the same server settings, increasing the number of service types necessarily worsen tail latency.
Therefore, in an edge distributed computation environment, tail latency is influenced by a combination of server performance and the types of services supported.
From the first subfigure of the Bottom row in Figure \ref{fig:tail_latency}, it is evident that, with a probability approaching $1$, the response time of the task is almost $20$ms, significantly smaller than the other subfigures.
This indicates that in the simulation experiment environment, the service types supported by each edge server are feasible for the distributed computation, and network resources are relatively abundant. 
% Although optimization of tail latency can yield better results, randomized and greedy algorithms show promising performance, with their \textit{CDF} curves closely resembling the \textit{CDF} curve of our proposed algorithm.
In general, our proposed algorithm guarantees enhanced robustness, adapting effectively to changes in the number of servers and service types.
%================================================================
% Queue Congestion
%================================================================
\subsubsection{Queue Congestion}
\begin{figure}[htbp]
        \includegraphics[width=0.32\linewidth]{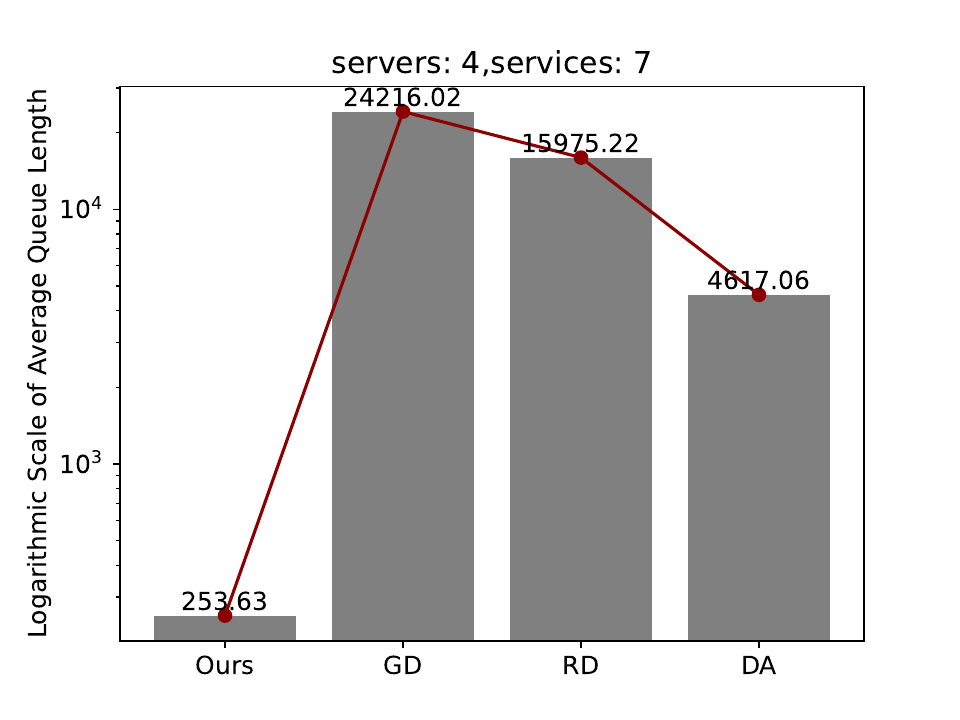}
        \includegraphics[width=0.32\linewidth]{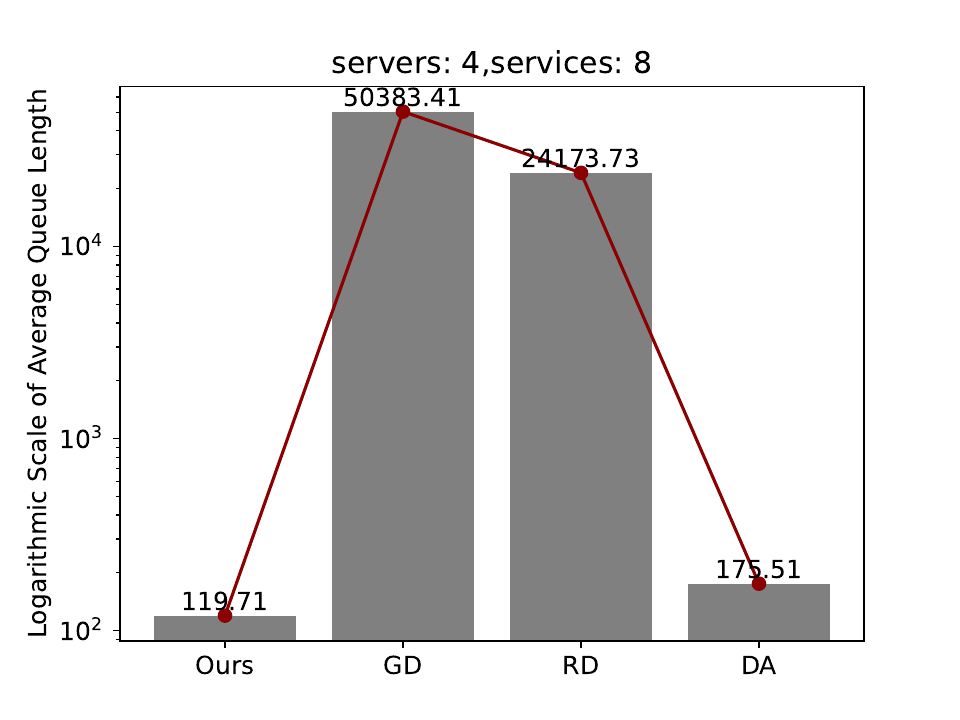}
        \includegraphics[width=0.32\linewidth]{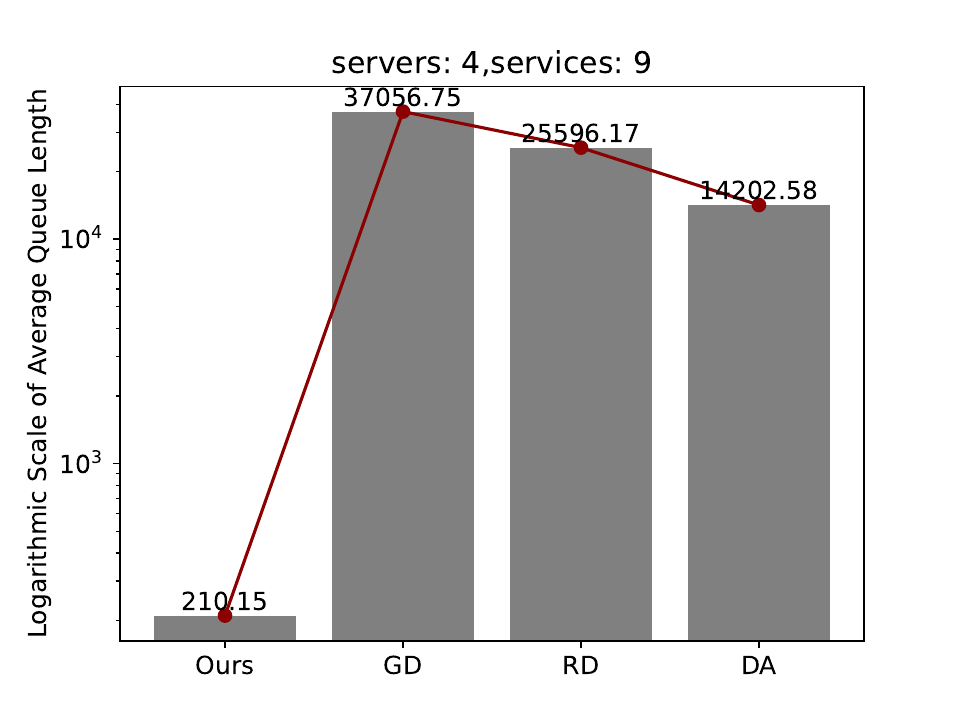}
        
        \includegraphics[width=0.32\linewidth]{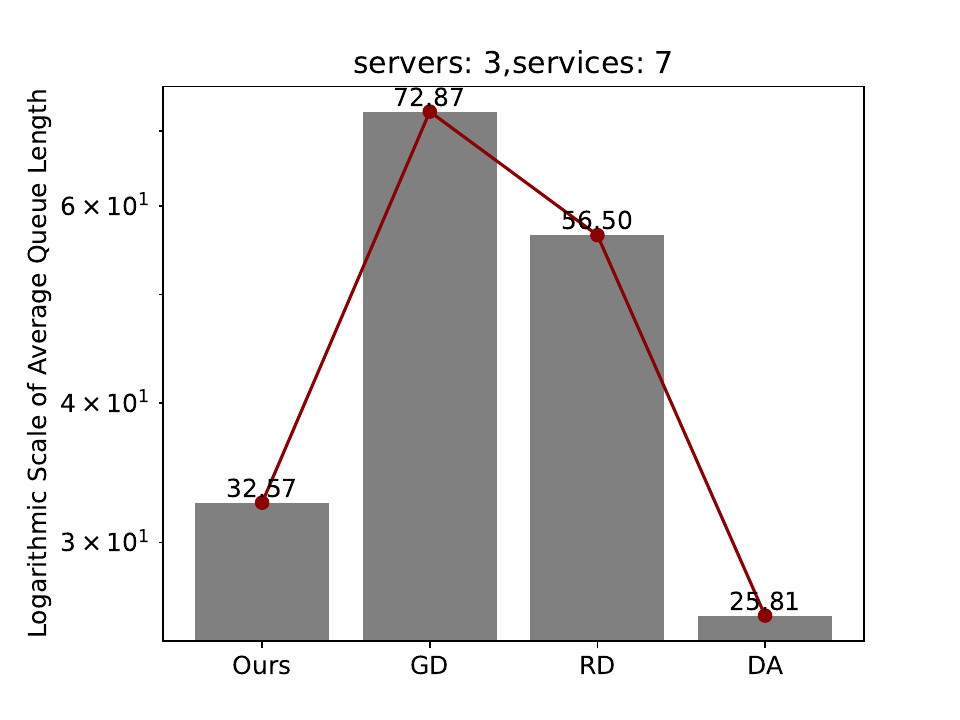}
        \includegraphics[width=0.32\linewidth]{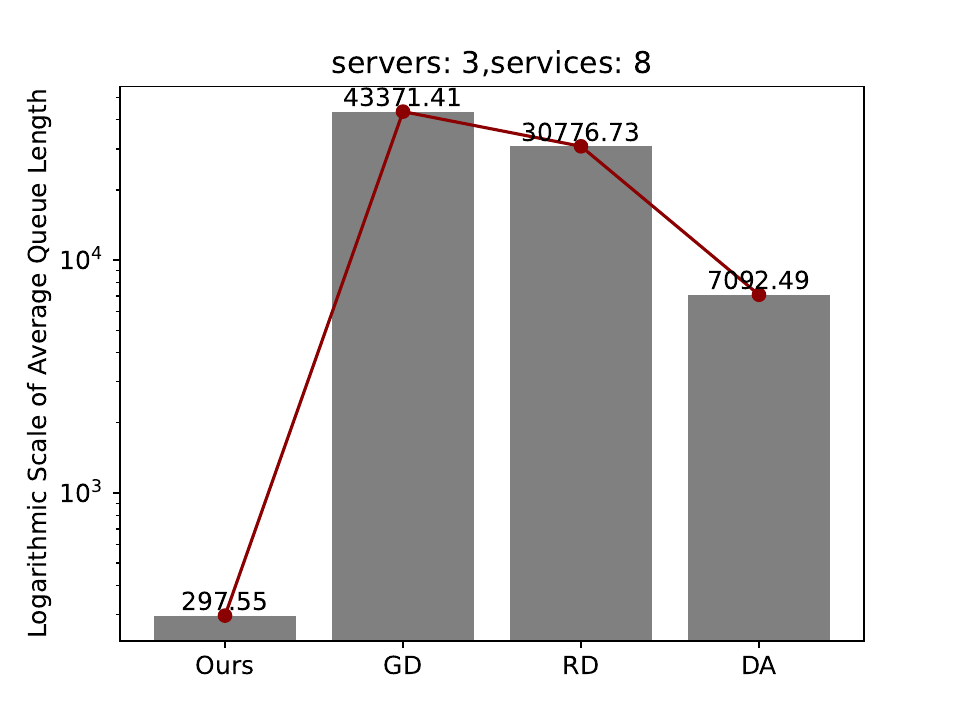}
        \includegraphics[width=0.32\linewidth]{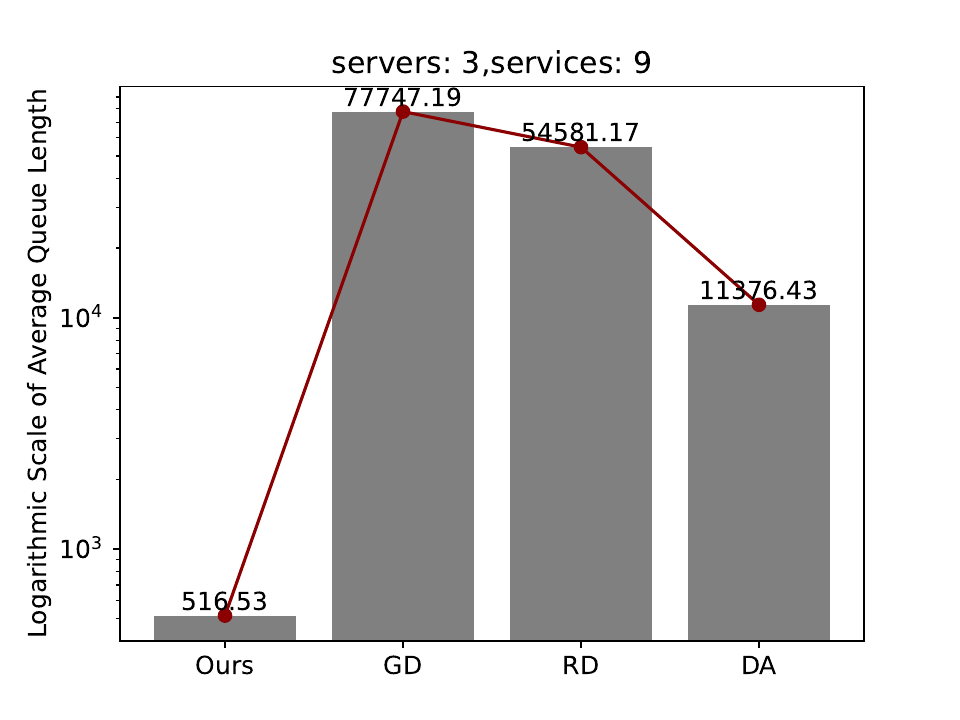}
    \caption{Average queue length for configurations with 3 and 4 servers under various service type combinations.}
    \label{fig:average_queue_leng}
\end{figure}
To generate Figure \ref{fig:average_queue_leng}, we accumulated the backlog of the three queues in the tandem queuing network presented in Figure \ref{fig_system}. 
The backlog values were recorded at each decision time, and the average backlog was computed for plotting in Figure \ref{fig:average_queue_leng}.

Figure \ref{fig:average_queue_leng} illustrates that, in most cases, the queue backlog of our proposed algorithm outperforms that of all compared algorithms, indicating reduced \textit{HoL} blocking.
The average queue length of the two comparative algorithms, \textit{GD} and \textit{RD}, is significantly larger than our proposed algorithm's.
Notably, the \textit{RD} and \textit{GD} algorithms, implemented without considering the negative impact of congestion, exhibit severe backlog issues. 
Additionally, when comparing the DA algorithm in both Figure \ref{fig:tail_latency} and Figure \ref{fig:average_queue_leng}, it is evident that the \textit{DA} algorithm shows a better tail delay response under conditions of light queue congestion, if compared to the \textit{RD} and \textit{GD} algorithms.
The definition of \textit{response time} as the sum of task \textit{waiting time} and \textit{execution time} highlights the significance of queue backlog in increasing waiting time and consequently aggravating tail latency.
Hence, in our reinforcement learning approach, the degree of queue congestion serves as a reward indicator, as detailed in Section \ref{sec_reward}. 
Although in the network environment with $3$ servers and $7$ types of services, where our queue backlog is slightly higher than that of \textit{DA}, Table \ref{tab:tail_metric} demonstrates that our tail latency performance remains superior to the \textit{DA} algorithm.
%================================================================
% Tail Latency Metric
%================================================================
\subsubsection{Performance under Percentiles}
% \usepackage{multirow}
% \usepackage[normalem]{ulem}
\begin{table}[]
\centering
\begin{tabular}{cccccccc}
\cline{1-7}
\multicolumn{1}{|c|}{\multirow{5}{*}{4-7}} & \multicolumn{1}{c|}{}                       & \multicolumn{1}{c|}{\textit{\textbf{\begin{tabular}[c]{@{}c@{}}p50\\ (Median)\end{tabular}}}} & \multicolumn{1}{c|}{\textit{\textbf{p90}}} & \multicolumn{1}{c|}{\textit{\textbf{p95}}} & \multicolumn{1}{c|}{\textit{\textbf{p99}}} & \multicolumn{1}{c|}{\textit{\textbf{p99.9}}} &                                        \\ \cline{2-7}
\multicolumn{1}{|c|}{}                     & \multicolumn{1}{c|}{\textit{\textbf{RD}}}   & 9.76                                                                                          & 5417.12                                    & 6062.86                                    & 6489.86                                    & \multicolumn{1}{c|}{6615.37}                 & \multirow{7}{*}{}                      \\ \cline{2-2}
\multicolumn{1}{|c|}{}                     & \multicolumn{1}{c|}{\textit{\textbf{GD}}}   & 2127.56                                                                                       & 6190.65                                    & 6720.97                                    & 7099.63                                    & \multicolumn{1}{c|}{7181.10}                 &                                        \\ \cline{2-2}
\multicolumn{1}{|c|}{}                     & \multicolumn{1}{c|}{\textit{\textbf{DA}}}   & 67.86                                                                                         & 1711.23                                    & 1903.55                                    & 2045.43                                    & \multicolumn{1}{c|}{2089.51}                 &                                        \\ \cline{2-2}
\multicolumn{1}{|c|}{}                     & \multicolumn{1}{c|}{\textit{\textbf{Ours}}} & {\ul \textit{5.76}}                                                                           & {\ul \textit{13.43}}                       & {\ul \textit{16.40}}                       & {\ul \textit{23.20}}                       & \multicolumn{1}{c|}{{\ul \textit{31.09}}}    &                                        \\ \cline{1-7}
\multicolumn{1}{|c|}{\multirow{5}{*}{4-8}} & \multicolumn{1}{c|}{}                       & \multicolumn{1}{c|}{\textit{\textbf{\begin{tabular}[c]{@{}c@{}}p50\\ (Median)\end{tabular}}}} & \multicolumn{1}{c|}{\textit{\textbf{p90}}} & \multicolumn{1}{c|}{\textit{\textbf{p95}}} & \multicolumn{1}{c|}{\textit{\textbf{p99}}} & \multicolumn{1}{c|}{\textit{\textbf{p99.9}}} &                                        \\ \cline{2-7}
\multicolumn{1}{|c|}{}                     & \multicolumn{1}{c|}{\textit{\textbf{RD}}}   & 5.84                                                                                          & 16926.32                                   & 20278.22                                   & 22847.62                                   & \multicolumn{1}{c|}{23125.45}                &                                        \\ \cline{2-2}
\multicolumn{1}{|c|}{}                     & \multicolumn{1}{c|}{\textit{\textbf{GD}}}   & 1571.76                                                                                       & 29549.20                                   & 33077.38                                   & 36135.78                                   & \multicolumn{1}{c|}{36562.94}                &                                        \\ \cline{2-2}
\multicolumn{1}{|c|}{}                     & \multicolumn{1}{c|}{\textit{\textbf{DA}}}   & 12.66                                                                                         & 184.75                                     & 277.73                                     & 452.39                                     & \multicolumn{1}{c|}{612.74}                  & \multirow{8}{*}{}                      \\ \cline{2-2}
\multicolumn{1}{|c|}{}                     & \multicolumn{1}{c|}{\textit{\textbf{Ours}}} & {\ul \textit{3.82}}                                                                           & {\ul \textit{8.92}}                        & {\ul \textit{11.05}}                       & {\ul \textit{16.13}}                       & \multicolumn{1}{c|}{{\ul \textit{23.47}}}    &                                        \\ \cline{1-7}
\multicolumn{1}{|c|}{\multirow{5}{*}{4-9}} & \multicolumn{1}{c|}{\textit{}}              & \multicolumn{1}{c|}{\textit{\textbf{\begin{tabular}[c]{@{}c@{}}p50\\ (Median)\end{tabular}}}} & \multicolumn{1}{c|}{\textit{\textbf{p90}}} & \multicolumn{1}{c|}{\textit{\textbf{p95}}} & \multicolumn{1}{c|}{\textit{\textbf{p99}}} & \multicolumn{1}{c|}{\textit{\textbf{p99.9}}} &                                        \\ \cline{2-7}
\multicolumn{1}{|c|}{}                     & \multicolumn{1}{c|}{\textit{\textbf{RD}}}   & 6.64                                                                                          & 8864.31                                    & 10356.89                                   & 11698.51                                   & \multicolumn{1}{c|}{11876.61}                &                                        \\ \cline{2-2}
\multicolumn{1}{|c|}{}                     & \multicolumn{1}{c|}{\textit{\textbf{GD}}}   & 1072.50                                                                                       & 10353.35                                   & 11459.84                                   & 12316.75                                   & \multicolumn{1}{c|}{12595.19}                &                                        \\ \cline{2-2}
\multicolumn{1}{|c|}{}                     & \multicolumn{1}{c|}{\textit{\textbf{DA}}}   & 28.34                                                                                         & 4669.46                                    & 5379.23                                    & 5977.62                                    & \multicolumn{1}{c|}{6138.78}                 &                                        \\ \cline{2-2}
\multicolumn{1}{|c|}{}                     & \multicolumn{1}{c|}{\textit{\textbf{Ours}}} & {\ul \textit{6.19}}                                                                           & {\ul \textit{45.80}}                       & {\ul \textit{69.61}}                       & {\ul \textit{112.62}}                      & \multicolumn{1}{c|}{{\ul \textit{139.43}}}   &                                        \\ \cline{1-7}
\multicolumn{1}{|c|}{\multirow{5}{*}{3-7}} & \multicolumn{1}{c|}{\textit{\textbf{}}}     & \multicolumn{1}{c|}{\textit{\textbf{\begin{tabular}[c]{@{}c@{}}p50\\ (Median)\end{tabular}}}} & \multicolumn{1}{c|}{\textit{\textbf{p90}}} & \multicolumn{1}{c|}{\textit{\textbf{p95}}} & \multicolumn{1}{c|}{\textit{\textbf{p99}}} & \multicolumn{1}{c|}{\textit{\textbf{p99.9}}} &                                        \\ \cline{2-7}
\multicolumn{1}{|c|}{}                     & \multicolumn{1}{c|}{\textit{\textbf{RD}}}   & 8.39                                                                                 & 47.21                       & 65.67                                      & 113.05                                     & \multicolumn{1}{c|}{165.92}                  & \multicolumn{1}{l}{\multirow{14}{*}{}} \\ \cline{2-2}
\multicolumn{1}{|c|}{}                     & \multicolumn{1}{c|}{\textit{\textbf{GD}}}   & 11.59                                                                                & 57.07                                      & 76.52                                      & 124.99                                     & \multicolumn{1}{c|}{155.60}                  & \multicolumn{1}{l}{}                   \\ \cline{2-2}
\multicolumn{1}{|c|}{}                     & \multicolumn{1}{c|}{\textit{\textbf{DA}}}   & 23.20                                                                               & 67.98                                     &85.50                                    & 128.45                                     & \multicolumn{1}{c|}{166.77}                  & \multicolumn{1}{l}{}                   \\ \cline{2-2}
\multicolumn{1}{|c|}{}                     & \multicolumn{1}{c|}{\textit{\textbf{Ours}}} & {\ul \textit{5.65}}                                                                           &{\ul \textit{12.15}}                                       & {\ul \textit{14.49}}                       & {\ul \textit{19.49}}                      & \multicolumn{1}{c|}{{\ul \textit{25.46}}}   & \multicolumn{1}{l}{}                   \\ \cline{1-7}
\multicolumn{1}{|c|}{\multirow{5}{*}{3-8}} & \multicolumn{1}{c|}{\textit{\textbf{}}}     & \multicolumn{1}{c|}{\textit{\textbf{\begin{tabular}[c]{@{}c@{}}p50\\ (Median)\end{tabular}}}} & \multicolumn{1}{c|}{\textit{\textbf{p90}}} & \multicolumn{1}{c|}{\textit{\textbf{p95}}} & \multicolumn{1}{c|}{\textit{\textbf{p99}}} & \multicolumn{1}{c|}{\textit{\textbf{p99.9}}} & \multicolumn{1}{l}{}                   \\ \cline{2-7}
\multicolumn{1}{|c|}{}                     & \multicolumn{1}{c|}{\textit{\textbf{RD}}}   & 8.40                                                                                 & 18561.31                                   & 21259.25                                   & 23515.21                                   & \multicolumn{1}{c|}{24049.01}                & \multicolumn{1}{l}{}                   \\ \cline{2-2}
\multicolumn{1}{|c|}{}                     & \multicolumn{1}{c|}{\textit{\textbf{GD}}}   & 4631.26                                                                             & 21424.84                                   & 23597.91                                   & 25302.42                                   & \multicolumn{1}{c|}{25764.06}                & \multicolumn{1}{l}{}                   \\ \cline{2-2}
\multicolumn{1}{|c|}{}                     & \multicolumn{1}{c|}{\textit{\textbf{DA}}}   & 9326.28                                                                             & 17455.60                                   & 19706.38                                   & 21557.77                                   & \multicolumn{1}{c|}{21938.69}                & \multicolumn{1}{l}{}                   \\ \cline{2-2}
\multicolumn{1}{|c|}{}                     & \multicolumn{1}{c|}{\textit{\textbf{Ours}}} & {\ul \textit{6.38}}                                                                           & {\ul \textit{16.21}}                       & {\ul \textit{21.35}}                       & {\ul \textit{35.08}}                       & \multicolumn{1}{c|}{{\ul \textit{49.46}}}    & \multicolumn{1}{l}{}                   \\ \cline{1-7}
\multicolumn{1}{|c|}{\multirow{5}{*}{3-9}} & \multicolumn{1}{c|}{\textit{\textbf{}}}     & \multicolumn{1}{c|}{\textit{\textbf{\begin{tabular}[c]{@{}c@{}}p50\\ (Median)\end{tabular}}}} & \multicolumn{1}{c|}{\textit{\textbf{p90}}} & \multicolumn{1}{c|}{\textit{\textbf{p95}}} & \multicolumn{1}{c|}{\textit{\textbf{p99}}} & \multicolumn{1}{c|}{\textit{\textbf{p99.9}}} & \multicolumn{1}{l}{}                   \\ \cline{2-7}
\multicolumn{1}{|c|}{}                     & \multicolumn{1}{c|}{\textit{\textbf{RD}}}   & {\ul \textit{9.94}}                                                                                & 19856.98                                   & 22978.54                                   & 25420.26                                   & \multicolumn{1}{c|}{25893.69}                & \multicolumn{1}{l}{}                   \\ \cline{2-2}
\multicolumn{1}{|c|}{}                     & \multicolumn{1}{c|}{\textit{\textbf{GD}}}   & 5561.36                                                                              & 23634.40                                   & 25892.06                                   & 27657.49                                   & \multicolumn{1}{c|}{28066.11}                & \multicolumn{1}{l}{}                   \\ \cline{2-2}
\multicolumn{1}{|c|}{}                     & \multicolumn{1}{c|}{\textit{\textbf{DA}}}   & 10986.50                                                                             & 20323.23                                   & 22569.35                                   & 24528.93                                   & \multicolumn{1}{c|}{24929.15}                & \multicolumn{1}{l}{}                   \\ \cline{2-2}
\multicolumn{1}{|c|}{}                     & \multicolumn{1}{c|}{\textit{\textbf{Ours}}} & {13.91}                                                                          & {\ul \textit{51.41}}                       & {\ul \textit{73.40}}                       & {\ul \textit{112.04}}                      & \multicolumn{1}{c|}{{\ul \textit{144.26}}}   & \multicolumn{1}{l}{}                   \\ \cline{1-7}
\multicolumn{7}{l}{}                                                                                                                                                                                                                                                                                                                                            & \multicolumn{1}{l}{}                  
\end{tabular}
\caption{Percentiles of distributed computation.}
\label{tab:tail_metric}
\end{table}
Table \ref{tab:tail_metric} presents a comprehensive overview of response times in the edge parallel computation system, utilizing metrics such as the median$(p50), p90, p95, p99,$ and $p99.9$.

Our proposed algorithm demonstrates significantly superior response times compared to the benchmark algorithm when assessed under the same metrics. 
As seen in Table \ref{tab:tail_metric}, when comparing the algorithms under the \textit{median} metric, the experimental results of our proposed algorithm are nearly in the same order of magnitude. However, once the percentile metric ranges from $p90$ to $p99.9$, the experimental results of the comparison algorithms far exceed the results of our proposed algorithm by several orders of magnitude. Therefore, the long-tail performance of our proposed algorithm is notably excellent, indicating minimal long-tail delay in cases close to the tail of tabular data.
In the configuration with 3 servers and 9 services, despite the median indicator suggesting a slightly larger response time for our algorithm than RD, our proposed algorithm significantly outperforms RD across $p90, p95, p99$, and $p99.9$ metrics. These improvements amount to $0.256\%, 0.319\%, 0.441\%$, and $0.557\%$ of the RD algorithm's response time, respectively, which underscores the efficacy of our proposed algorithm, particularly in mitigating higher percentiles of response times.
%================================================================
%  Enhanced State of Tail-learning
%================================================================
\subsubsection{Enhanced State of Tail-learning}
\begin{figure}
\centering
    \includegraphics[width=.55\linewidth]{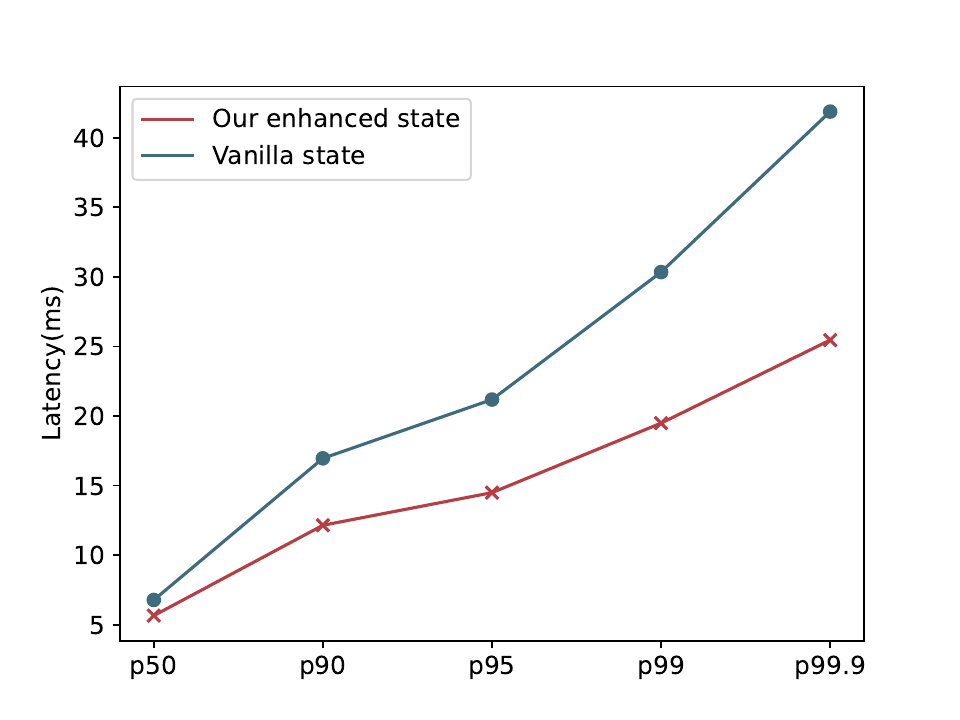}%
    \caption{Percentiles performance under $p50$, $p90$, $p99$ and $p99.9$}
    \label{fig:rl_latency_constrast}
\end{figure}
\begin{figure}
    \centering
    \includegraphics[width=.55\linewidth]{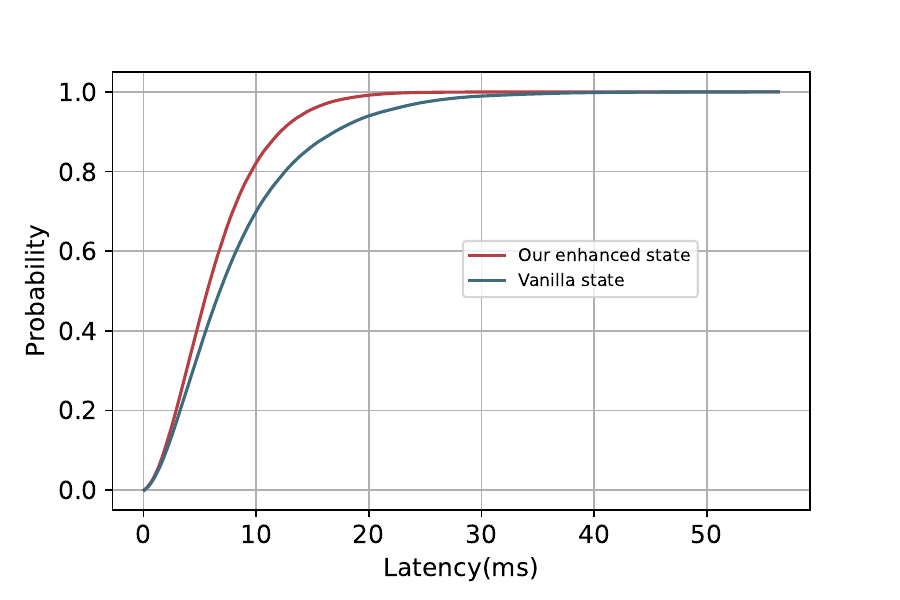}
    \caption{CDFs for \textit{QLA} and our proposed method}
    \label{fig:rl_cdf_constrast}
\end{figure}
Utilizing a tandem queueing system, the \textit{QLA} method focuses on optimizing end-to-end delay, considering queue lengths as the state in a policy-based method. 
We adopt \textit{QLA} \cite{raeis2021queue} to illustrate that our designed state in Section \ref{sec_enhanced_state} can enhance the adaptive learning with a high performance. 
Our proposed algorithm shares similarities with \textit{QLA}, as both algorithms track whether the response time of a request exceeds a threshold within a specified time range $\Delta$ (refer to Section \ref{sec_reward}) and subsequently compute the reward for each step. 
The key distinction lies in the fact that the \textit{QLA} algorithm solely utilizes the queue length as the state variable.
However, our algorithm goes beyond \textit{QLA} by extending the vanilla state, incorporating information related to tail latency probability. 
We use the enhanced state to represent this construction method, analyzing and proving theoretical bounds on tail latency probabilities in Section \ref{sec_tail_bound}. 
We applied the \textit{QLA} principle within our proposed algorithm framework in the evaluation. 
Experiments were conducted with our proposed and \textit{QLA} algorithms on $3$ servers, testing $7$ service types with identical parameters used in Figure \ref{fig:tail_latency}.
\begin{figure}
\centering
        \includegraphics[width=0.55\linewidth]{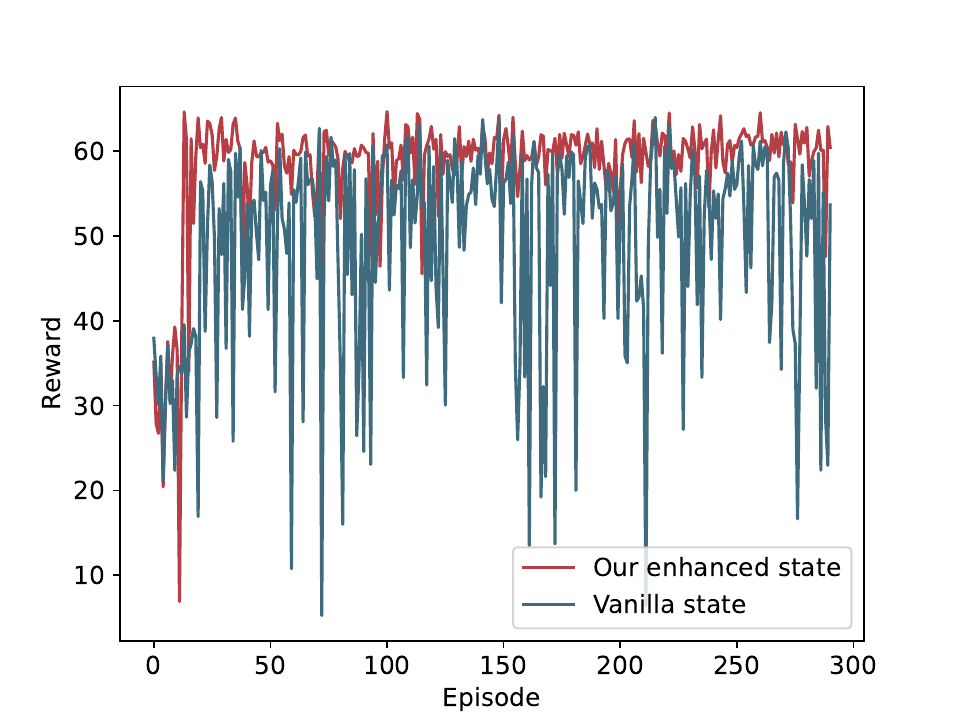}
        \caption{Rewards plotting for \textit{QLA} and our proposed method}
        \label{fig:rl_reward_constrast}
\end{figure}
Figure \ref{fig:rl_cdf_constrast}'s CDF curves illustrate that our proposed algorithm ensures a higher probability of shorter response times than \textit{QLA}. 
Additionally, each percentile point in Figure \ref{fig:rl_latency_constrast} indicates that the enhanced state for \textit{tail-learning} significantly mitigates tail delays, reaching followed a descending order only $83.41\%, 71.63\%, 68.41\%, 64.22\%$ and $60.78\%$ of \textit{QLA}. 
The experimental results indicate that our state-enhanced learning strategy performs better as the tail metric ($p50$, $p90$, $p99$, and $p99.9$) increases.
The reward curve in Figure \ref{fig:rl_reward_constrast} demonstrates that our proposed enhanced state method achieves higher rewards than \textit{QLA}. 
The curves consistently outperform the \textit{QLA} reward curve, exhibiting more excellent stability and lower volatility. 
This observation suggests better adaptive performance to the environment and significant improvement in the reinforcement learning process.
Furthermore, Figure \ref{fig:rl_reward_constrast} demonstrates that the enhanced state we propose exhibits a more stable convergence effect than the vanilla state in \textit{QLA}. 
Therefore,  our enhanced state design is practical and efficient.
%================================================================
% Conclusions
%================================================================
\section{Conclusions and Future Work}\label{sec_conculsion}
In the dynamic field of edge computing, maintaining high QoS is paramount. However, the shift to a distributed edge paradigm has heightened the challenge of tail latency. Traditional queuing exacerbates issues like head-of-line blocking.
To address this, we present \textit{Tail-learning}, a novel scheduling method that dynamically selects edge servers. We achieve significant reductions in tail latency by leveraging Laplace Transform techniques and incorporating them into a reinforcement learning framework, as demonstrated in simulations.
Our contributions involve formulating response time dynamics, introducing \textit{Tail-learning}, and showcasing its superiority through experimentation. 
In future research, prioritizing services becomes essential, mainly when service resources are limited, leading to significantly lower long-tail probabilities for some services than others. 
Concurrently, researchers should focus on comprehending task dependencies and explore practical approaches to handle long-tail optimization associated with service dependencies.

% \begin{acks}
% \end{acks}

\bibliographystyle{ACM-Reference-Format}
\bibliography{sample-base}

% \appendix

\end{document}